\documentclass[reprint,superscriptaddress,groupedaddress,
 amsmath,amssymb,
 aps,
prb,
floatfix,
]{revtex4-2}

\usepackage{graphicx}% Include figure files
\usepackage{dcolumn}% Align table columns on decimal point
\usepackage{bm}% bold math
\usepackage{graphicx}
\usepackage[normalem]{ulem}
\usepackage{bm}
\usepackage{amsmath}
\usepackage{amssymb}
\usepackage{braket}
\usepackage{xcolor}
\begin{document}

\title{Optical band engineering via vertical stacking of honeycomb plasmonic lattices}
\author{D. Becerril, G. Pirruccio and Cecilia Noguez}
\email[Corresponding author: ]{cecilia@fisica.unam.mx}
\affiliation{Instituto de F\'{i}sica, Universidad Nacional Aut\'{o}noma de M\'{e}xico, Ciudad de M\'{e}xico C.P. 04510,   Mexico}

\date{\today}

\begin{abstract}
Inspired by recent advances in atomic homo and heterostructures, we consider the vertical stacking of plasmonic lattices as a new degree of freedom to create a coupled system showing a modified optical response concerning the monolayer. The precise design of the stacking and the geometrical parameters of two honeycomb plasmonic lattices tailors the interaction among their metallic nanoparticles. Based on the similarity of the lattice symmetry, analogies can be drawn with stacked atomic crystals, such as graphene. We use the multipolar spectral representation to study the plasmonic vertical stack's optical response in the near-field regime, emphasizing symmetry properties. The strong coupling of certain optical bands and polarization switch of the interacting bands. By leveraging these effects, we engineer the near-field intensity distribution. Besides, lifting band degeneracy at specific points of the Brillouin zone is obtained with the consequent opening of mini-gaps. These effects are understood by quantifying the multipolar coupling among nanospheres belonging to the same and different sublattices, as well as the interlayer and intralayer nanoparticle interactions. Differences with the atomic case are also analyzed and explained in terms of the stack's interaction matrix. Finally, we predict the absorption spectrum projected on the two orthogonal linear polarizations.
\end{abstract}

\maketitle

\section{Introduction}

Among the possible lattice symmetries, the honeycomb one has gathered enormous attention after the successful synthesis of graphene, which signaled the birth of genuinely two-dimensional (2D) physics in atomic systems. Many of the fascinating graphene properties relate to its bipartite, non-Bravais crystalline structure, the most outstanding example being the Dirac cones with their linear dispersion. Remarkable is also the breaking of the valley degeneracy at the K-points of the Brillouin zone. It is currently being exploited to tune the optoelectronic properties of new materials and sparked the investigation of a variety of valley-sensitive phenomena. 

Recently, the relative orientation of 2D atomic vertically-staked lattices has been proposed as a new geometrical degree of freedom to engineer their electronic band structure. The relative orientation, or twist angle, of stacked graphene and dichalcogenides layers has been shown to influence their electronic, optical, chemical, and mechanical properties. Bilayers of 2D layered materials exhibits a wide variety of novel physical properties, manifested as profound modifications of the electronic structure \cite{Bistritzer12233}, interlayer coupling \cite{Liu,doi:10.1021/acsnano.5b01884}, optical \cite{PhysRevB.87.205404}, and insulating properties \cite{Cao20182}. Special twist angles, termed as magic angles, are responsible for the emergence of spectacular effects such as superconduction \cite{Cao2018}, topological phases \cite{PhysRevLett.122.086402} and spin waves \cite{PhysRevB.100.035413}. 

After reckoning the properties of the honeycomb lattice, many different analogies have been realized, aiming at replicating the physics of graphene and 2D materials at different scales. Artificial graphene systems differ from the atomic case because of the type of interaction within the lattice. One of the closest optical analogies of 2D materials is 2D plasmonic lattices made of resonant metallic nanoparticles, which coupled together via an external electromagnetic field, whereby the electronic bands are replaced by optical bands, and collective plasmon modes \cite{artgraph}.

The electromagnetic interaction in plasmonic lattices can be divided into two distinct regimes: the diffractive regime and the near-field regime. In the diffractive regime, the interparticle distance is on the order of the incident wavelength, and thus diffraction can favor the radiative coupling between the localized surface plasmon resonances associated with each isolated nanoparticle \cite{chemrev}. Under this special condition, the 2D plasmonic lattice displays an optical band structure whose Bloch modes are termed as Surface Lattice Resonances (SLRs) \cite{RevModPhys.79.1267,PhysRevLett.116.103002,WANG2018303}. These modes show a collective behavior on the long-range scale, involving several unit cells. They have been recently studied also in non-Bravais \cite{D0NA00095G,doi:10.1021/acsnano.7b08206,doi:10.1021/acsnano.0c04795} and vertically stacked plasmonic lattices \cite{PhysRevApplied.14.054030}. Diffractive lattices present striking spectral far-field properties, among which Fano-like resonances, electromagnetically induced transparency windows, ultra-sharp linewidth are just few examples \cite{PhysRevB.80.201401,doi:10.1021/jz2002452,PhysRevLett.101.143902,doi:10.1021/ph400072z}. 

In analogy with graphene, non-trivial phenomena occur mainly at the K-points of the first Brillouin zone of the honeycomb plasmonic lattice \cite{PhysRevLett.122.013901}. This type of lattice can present a more involved response than a Bravais since its base can be used to engineer its optical properties \cite{Ribeiro-Palau690}. Using a dipole approximation, it has been shown that Dirac cones, similar to the $\pi$ bands of electrons in graphene, exist at the K-points of honeycomb plasmonic lattices. Besides, Dirac cones have already demonstrated to host intriguing effects related to parity-time symmetry, and non-hermitian physics \cite{Mirieaar7709}. The presence of exceptional points in the dispersion diagram of 2D lattices with balanced gain and loss provides the honeycomb lattice with a non-trivial topological optical band structure \cite{8852732}.

In this work, we use the multipolar spectral representation (MSR) to study a honeycomb monolayer of plasmonic nanoparticles, first, and then a vertically stacked bilayer. By focusing on the A-A and A-B stacking, we stress similarities and differences between the atomic crystal and the optical lattice hamiltonians. Near-field multipolar interaction between nanoparticles is found to be a direct electromagnetic analogy of the covalent interactions between atomic orbitals. It is well known that in a two-dimensional plasmonic honeycomb monolayer described by point dipoles, the in-plane modes couple together while they remain decoupled from the out-of-plane ones \cite{PhysRevLett.110.106801}. We find that in the coupled bilayer, the subwavelength distance between the individual layers results in interlayer band interactions sufficiently strong to cause level repulsion at the crossing of orthogonally polarized bands and the opening of mini-gaps at the K-point of the first Brillouin zone. These properties are revealed both in the calculated absorption spectrum and spatial near-field distribution.

\section{Theoretical Model and Method}

We consider a 2D honeycomb lattice composed of non-magnetic polarizable nanospheres of radii $a$  whose response to an external potential is described by a frequency-dependent local dielectric function $\epsilon(\omega)$, and are within a host matrix with dielectric constant $\epsilon_{h}$. 
The nanospheres are sufficiently close to each other, so they near-field couple through Coulomb interactions between the induced charges on each particle. The local field felt by each particle, which results from summing up the contributions from all the nanospheres and the external field, gives rise to a multipolar charge distribution on each sphere. By virtue of the lattice translational invariance, the associated induced multipole moments are written as Bloch functions, which imply a fixed phase relation between the spheres. 

To describe the response of the system to external excitations, we use the MSR \cite{PhysRevB.34.3730,PhysRevB.57.302}, which allows a systematic analysis of the system eigenmodes and its dependence on multipolar interactions \cite{doi:10.1021/acsphotonics.7b01426}. This method relies on the multipolar expansion of the electrostatic potential to obtain a set of equations for the induced multipolar charge distribution on each of the particles in the system. A  more detailed overview is given in the Appendix. Using a ket notation \cite{8852732,PhysRevLett.91.227402,PhysRevB.22.4950} for the induced multipolar moments $\ket{X}$ and the external potential $\ket{F(\vec{k})}$ acting on each particle of the system, we obtain a set of equations 
\begin{equation}\label{secular}
\left[-u(\omega)\mathbf{I}+\mathbf{H}(\vec{k}) \right]\ket{X}=\ket{F(\vec{k})},
\end{equation}
where $u(\omega)=1/[1-\epsilon(\omega)/\epsilon_{h}]$ is known as the complex spectral variable, $\mathbf{I}$ is the unitary operator, and $\mathbf{H}$ is a hermitian operator, which describes the interaction between multipolar charge distributions and depends only on the geometrical parameters of the system. Since we are working with spherical scatters, it is convenient to  expand in a basis of spherical harmonics $\ket{\phi_\mu}$ centered at each particle, analogous to the expansion of the crystal wavefunction as a linear combination of localized atomic orbitals \cite{grosso2000solid}. 

On this basis, the components of kets in Eq.~\eqref{secular} are given as
\begin{eqnarray}\label{eq:TBSRHelement}
\braket{\phi_\mu| X } &=& \frac{Q_{\mu}}{\sqrt{l a_{i}^{2l+1}}}\, ,\label{X}\\
\braket{\phi_\mu | F(\vec{k}) }&=& -\frac{\sqrt{l a_{i}^{2l+1}}}{4\pi}V^{\text{ext}}_\mu e^{j \vec{R}_{i} \cdot \vec{k}}  \,,\label{Fa} \hspace{ 1cm}  
\end{eqnarray}
where $\mu$ is shorthand for the set of indexes $(l,m,i)$. Term $Q_\mu$ is the $lm$-th spherical multipole on the $i$-th particle in the unit cell, $V^{\text{ext}}_{\mu}$ is the $lm$-th term of the expansion of the external potential at $i$, $j$ is the imaginary unit,  and $R_{i}$ is the coordinate of particle $i$ in the unit cell.

Solutions to Eq.~\eqref{eq:TBSRHelement} can be found by constructing the  Green's function, $\mathbf{G}= -\sum_{s} \frac{\ket{s}\bra{s}}{u(\omega)-n_{s}}$ \cite{PhysRevB.70.195412}, where $\ket{s}$ are the system eigenvectors obtained by solving the eigenvalue equation
\begin{equation}\label{eq:eigen}
\mathbf{H}\ket{s} = n_s\ket{s}. 
\end{equation} 
where $\mathbf{H}$ has components  given by Eq.~\eqref{eq:Hmat}, shown in the Appendix. We point out that $\mathbf{H}$ can be split into two matrices, one related to the response of the isolated nanosphere and the other to a periodic matrix that describes the interaction potential within the lattice. This equation is akin to the Schroedinger equation found in solid state physics for the crystal hamiltonians and its solution describes the formation of optical bands in plasmonic lattices in the near-field regime. 

Once solved using the spherical harmonic basis $\ket{\phi_\mu}$, the eigenstate components reads $\braket{\phi_\mu| s} = U_{\phi_\mu s}$, where $U_{\mu s}$  is a unitary matrix as defined in the Appendix. The product of two eigenstates is simply $\braket{s| s^\prime} = \sum_{\mu} U^{*}_{s \mu} U_{\mu s^{\prime}} = \delta_{s s^\prime}$ consistent with the  orthogonality of eigenstates.
This eigenvalue equation clarifies the analogy between the atomic and optical lattices, whereby the hamiltonian and electronic bands are replaced by the $\mathbf{H}$ matrix and optical bands, respectively. The electron interactions within the atomic lattice are here replaced by the Coulomb interaction between charge densities.

Using a complete orthonormal basis, the Green matrix has components:
\begin{equation}\label{green0}
\bra{\phi_\mu}\mathbf{G}\ket{\phi_{\mu^\prime}} = -\sum_{s} \frac{\braket{\phi_\mu|s}\braket{s|\phi_{\mu^\prime}}}{u(\omega)-n_{s}} = -\sum_{s} \frac{\text{C}_{\mu}^{\mu^{\prime}}(s) }{u(\omega)-n_{s}},
\end{equation}
where $n_s$ are the eigenvalues of matrix $\mathbf{H}$ associated with the modes, $s$, of the system. $\text{C}_{\mu}^{\mu^{\prime} }(s)$ are components of matrix $\mathbf{C}(s)$, which describe the coupling strength of the external fields through the s-th mode of the system (see the Appendix).   Note that the eigenvalues, or equivalently the modes of the system, are independent of the external potential and only depend on the geometrical parameters of the system, see Eq.~\eqref{eq:eigen}. In other words, the band structure of the plasmonic lattice is fixed by the symmetry of the lattice, while the spectral range where the bands are found is material-dependent. 

The denominator of the Green matrix defines resonance conditions given as:
\begin{equation}\label{eq:res_condition}
	\text{Re}[u(\omega)] = n_s(\vec{k}), \hspace{1cm} \text{Im}[u(\omega)] \ll 1
\end{equation}
To find the resonant energy,  $n_s(\vec{k})$, of a given mode, it is necessary to define a dielectric function. For example, when using a Drude dielectric function of the form
\[\epsilon(\omega)  = 1 - \frac{\omega_p^2}{\omega(\omega + j\tau^{-1}+ j\tau(a)^{-1} )},\] with $\tau^{-1}$ and $\tau(a)^{-1}$ being the usual scattering rate and the one corrected for the small nanoparticle dimension, respectively. Thus the frequency, $\omega_s(\vec{k})$, of the s-th eigenmode is given by \cite{doi:10.1021/jp066539m}
\begin{equation}\label{eq:ResFrequDrude}
\omega_s(\vec{k}) = -j \Gamma + \sqrt{\omega_p^2 n_s(\vec{k}) A_s(\vec{k}) - \Gamma^2 }
\end{equation}
where $A_s(\vec{k})  = [n_s(\vec{k})(1-\epsilon_h) + \epsilon_h ]^{-1}$ and $\Gamma = \frac{1}{2}(\tau^{-1}+\tau(a)^{-1} )$\cite{doi:10.1021/jp066539m}. Eq.~\eqref{eq:ResFrequDrude} defines the dispersion relation for the modes of a lattice of spherical NPs described by a Drude dielectric function.

To calculate the induced multipolar moments on each particle, we use the Green function:
\begin{equation}\label{eq:Sol1}
\ket{X} = \mathbf{G}\ket{F}\,.
\end{equation}
Once the induced moments are obtained, physical properties such as the the absorption cross section $C_{abs}(\omega)$ can be calculated. For example, $C_{abs}(\omega)$ is proportional to the dipole moments, $\vec{p}=(p_x,p_y,p_z)$, of the system, so using Eq.~\eqref{eq:Sol1} and the relations
\begin{equation}\label{eq:dipoles}
\begin{split}
Q_{l = 1, m = 0} & = \sqrt{\frac{3}{4\pi}} p_z \\
Q_{l = 1, m \pm 1} & =- \sqrt{\frac{3}{8\pi}} \left(p_x  \mp j p_y\right) \\
\end{split}
\end{equation}
we can calculate the absorption cross section for a given external field
\begin{equation}\label{eq:cabs}
C_{abs} = 4\pi \left( \frac{2\pi\sqrt{\epsilon_h}}{\lambda} \right) \sum_{i}^{N_p} \text{Im}\left[ \vec{E}_{ext}\cdot \vec{p}_i\right]
\end{equation}
where $\lambda$ is the vacuum wavelength of the external field.  Notice that  dipoles in Eqs.~\eqref{eq:dipoles}--\eqref{eq:cabs} are calculated using Eq.~\eqref{eq:Sol1}. Therefore all multipolar interaction  contribute to the induced dipole moments, which is described through the Green matrix. In principle, all the matrices in the previous equations are infinite-dimensional. 

To correctly solve the equations,  we introduce the parameter $l_{max}$, which defines the maximum multipole moment considered and therefore determines the dimension of matrices $\mathbf{H}$, $\mathbf{G}$, etc. In general, for a given
set of geometrical parameters, such as separation distance, particle radius, and lattice symmetry, an appropriate $l_{max}$ must be chosen to ensure convergence of the physical
properties. For $l_{max} = 1$ we recover the dipole approximation, which is expected to be good for large particle separations \cite{Wang_2016,8852732}. For $l_{max} = 2$
the quadrupole approximation is obtained in which dipole-dipole, dipole-quadrupole and
quadrupole-quadrupole moments interactions are taken into account \cite{doi:10.1021/acsphotonics.7b01426}. 

 \begin{figure}[b]
	\includegraphics[width= 0.45\textwidth]{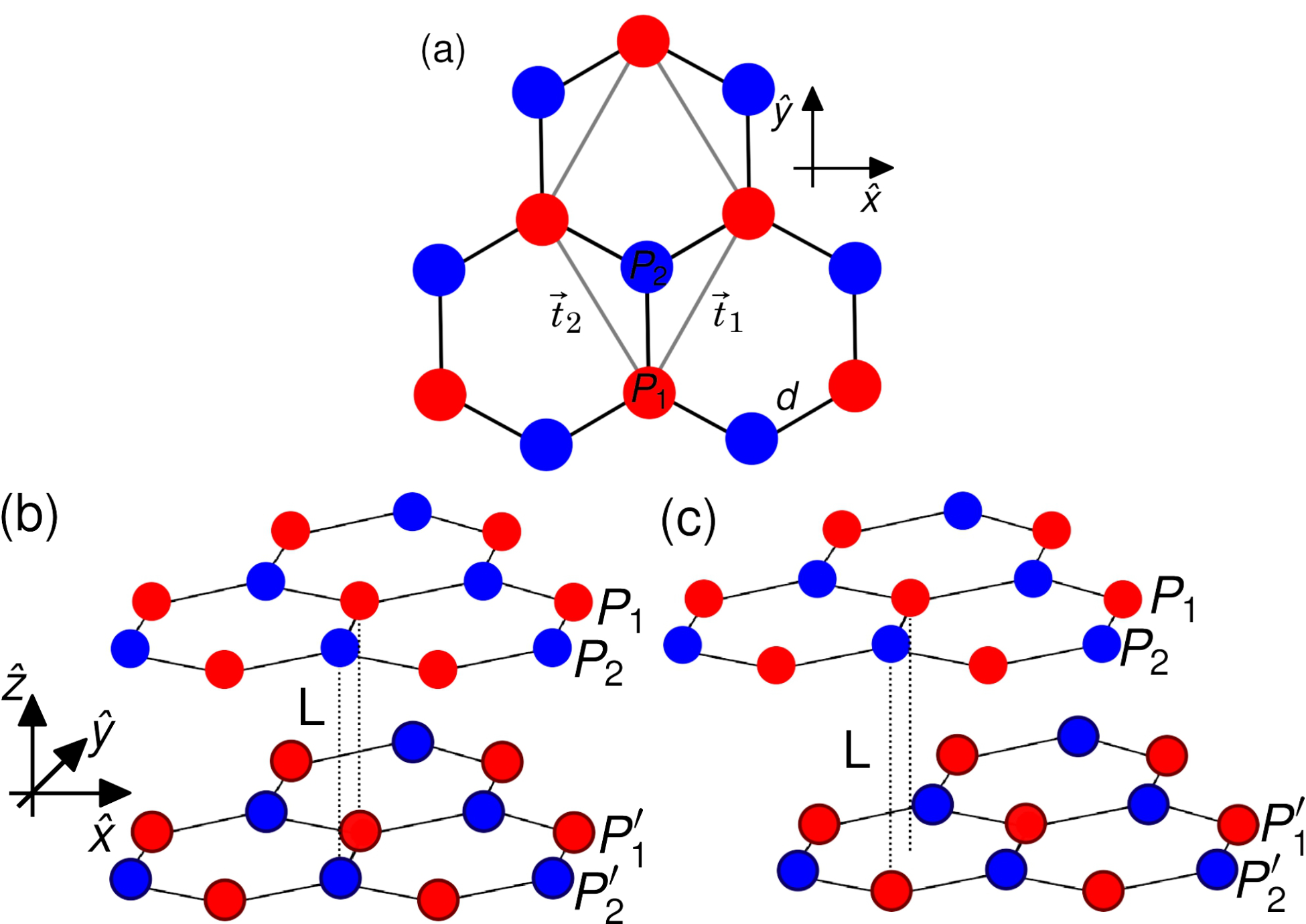}
	\caption{(a) Schematic of the monolayer  with two particles in each unit cell. Nearest neighbor vectors, $\vec{d}_i$, and translation vectors, $\vec{t}_1$ and $\vec{t}_2$, are shown. (b) A-A and (c) A-B vertical stacking with interlayer distance $L$. Unit cell now has 4 particles, two per layer denoted as $P_1,P_2$ and $P^\prime_1,P_2^\prime$.    }
	\label{fig:1}
\end{figure} 

\begin{figure*}
  	\includegraphics[width= 1.00\textwidth]{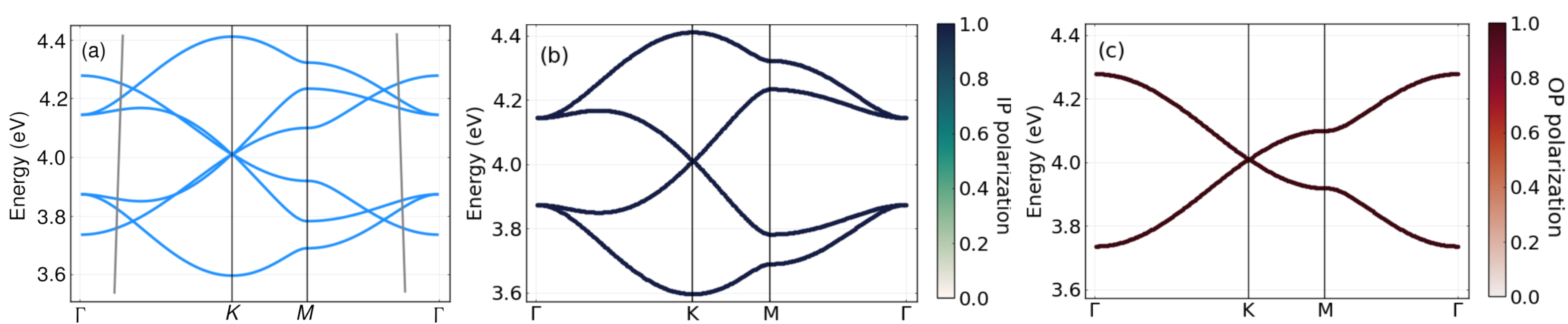} 
 	\caption{(a) Dispersion relation  along the $\Gamma - K-M$ path of the reciprocal space for the honeycomb monolayer with the light line shown in gray, using Eq.~\eqref{eq:ResFrequDrude}. (b) In-plane $P_{IO}(s)$,  and (c) out-of-plane $P_{OP}(s)$  polarization calculated using Eq.~\eqref{eq:polarization0} at each mode $s$. }
 	\label{fig:2}
 \end{figure*}

It will  also be useful to calculate other physical properties, such as the polarization of a given mode. For example, it is convenient to introduce a quantity that measures the in-plane (IP) and out-of-plane (OP) character of a given mode.  To accomplish this, we can project a given mode onto the part of the basis with IP  and OP  dipolar symmetry.  In the case of a spherical harmonic basis this  corresponds to projecting onto basis functions  with index $l = 1, m = 0$ (out-of-plane) and $l = 1, m\pm1$ (in-plane). In the bracket notation, we define the in-plane (IP) and out-of-plane (OP) polarization of a  mode as
\begin{equation}\label{eq:polarization0}
\begin{split}
P_{IP} &= \bra{s} \left( \sum_{\substack{ l=1\\m \pm  1}}\ket{\phi_\mu}\bra{\phi_\mu} \right) \ket{s} =  \sum_{\substack{ l=1\\m \pm  1}}C^{\mu}_{\mu}(s) \\
P_{OP} & = \bra{s}\left( \sum_{\substack{ l=1\\m =0}}\ket{\phi_\mu}\bra{\phi_\mu} \right) \ket{s} = \sum_{\substack{ l=1\\m =0}}C^{\mu}_{\mu}(s)
\end{split}
\end{equation}
Using Eq.~\eqref{green0}, it can be verified that this  quantity is equivalent to taking the sum of coupling weights of a given mode to  an  IP (OP) external field. It has been shown that coupling strengths of the form $C^{\mu}_{\mu}(s) $,   take only positive values and that the trace of coupling matrix $\mathbf{C}(s)$ satisfies $ \sum_{\mu}  C_{\mu}^{\mu}(s) = 1$, so that the polarization of a mode defined by Eq.~\eqref{eq:polarization0} takes values between 0 and 1 \cite{PhysRevB.34.3730}. This model can be extended to an arbitrary number of vertically stacked plasmonic lattices by constructing the interaction matrix H with blocks corresponding to the inter and intra layer particle interaction and whose dimensionality is determined by $l_{max}$, as shown in the Appendix.

\section{Results and Discussion}
\subsection*{Monolayer}

We begin by applying our method to the case of a honeycomb monolayer of $Ag$ nanospheres  with radius $a = 10$ nm, described by a Drude dielectric function,  and  hosted within a homogeneous medium of dielectric constant $\sqrt{\epsilon_h} = 1.46$, which describe  the substrate used in experimental systems \cite{D0NA00095G}. The  separation distance between particle centers in a layer is $d = 3a$, so that we can safely restrict our calculations to the dipole approximation  ($l_{max} = 1$) \cite{doi:10.1021/acsphotonics.7b01426}. This system has been extensively studied due to its analogy with 2-D electronic systems \cite{PhysRevLett.110.106801}.

For a strictly 2-D configuration adequately described by the dipole approximation, the interaction between orthogonally oriented multipolar moments  is prohibited, as shown in the Appendix. Because of this, IP and OP modes can be solved independently, which corresponds to solving indexes $m = \pm 1$ and $m = 0$ separately. The dispersion relation $n_s(\vec{k})$ for  OP modes is particularly simple and can be written as:
\begin{equation}\label{eq:dispersion0}
	n_{OP}(\vec{k}) = n_{0,1} \pm \left|\mathbf{H}_{m = 0, i = 1}^{m^\prime = 0, i^\prime = 2}(\vec{k})\right| = n_{0,1} \pm \frac{2a^3}{\sqrt{3}} \frac{1}{d^3} \left| F(\vec{k}) \right|,
\end{equation} 
 here $\mathbf{H}_{m = 0,1}^{m = 0,1}(\vec{k})$ is the OP part of the interaction matrix between particles $P_1$ and $P_2$ as shown in schematic of Fig.~\ref{fig:1}; $n_{0,1}= 1/3$, is the eigenvalue of an isolated dipole, and $F(\vec{k}) = \sum_{i}^{3}e^{j \vec{k}\cdot\vec{d_i}} $ where $\vec{d}_i = d\hat{d}_i$ is one of three nearest-neighbor vectors as shown in schematic of Fig.~\ref{fig:1}. 

 Notice how Eq.~\eqref{eq:dispersion0} only depends on components of the interaction matrix with $m = m^\prime = 0$. This equation is analogous to the dispersion relation obtained using a tight-binding model for the electronic bands in graphene, where the term $\left|\mathbf{H}_{m = 0, i = 1}^{m^\prime = 0, i^\prime = 2}(\vec{k})\right|$, is analogous to the hopping integrals between nearest neighbor orbitals \cite{grosso2000solid}. Eq.~\eqref{eq:dispersion0} illustrates how 
the system's eigenvalues deviate from that of an isolated sphere $n_{0,1}$ due to the contributions from  dipole-dipole interaction described by $\mathbf{H}_{m = 0, i = 1}^{m^\prime = 0, i^\prime = 2}(\vec{k})$ and  $F(\vec{k})$. 
From Eq.~\eqref{eq:dispersion0} we can also see that the band width decreases  as $1/d^3$ as nearest-neighbor separation increases.  An equivalent, though slightly more complicated relation may be obtained for the IP eigenvalues written in terms of $H^{1}_{1}$, $H^{-1}_{-1}$, $H^{1}_{-1}$, and $H^{-1}_{1}$.

 The complete band structure for the honeycomb monolayer is calculated by solving Eq.~\eqref{eq:eigen} along the high-symmetry trajectories and using Eq.~\eqref{eq:ResFrequDrude} as is shown in Fig.~\ref{fig:2}(a).  In accordance with past work \cite{Wang_2016}, there is a Dirac-like cone  at the $K$-point. Using Eq.~\eqref{eq:dispersion0} it can be seen that, similar to the case of graphene, the dispersion relation for OP modes is linear in $k$  near the  K-point.  Note that at the K-point, the interacting term in Eq.~\eqref{eq:dispersion0} goes to zeros, and the eigenvalue at  K-point is degenerate and equal to that of the isolated particle.   In contrast to the case of graphene, the IP modes also form a Dirac cone near the $K$-point. This difference is due to the fact that IP modes in graphene are formed by hybridized s-p orbitals, while NPs interact only via dipole fields with an equal symmetry to those of p-orbitals. 

To illustrate which modes are IP or OP, Fig.~\ref{fig:2}(b) and (c) show the mode polarization along the high-symmetry trajectory $\Gamma-K-M$ calculated using Eq.~\eqref{eq:polarization0}. In agreement with previous work, as well as with the condition prohibiting IP and OP interaction, it can be seen that modes are either completely IP or OP. As pointed out in the previous section, the polarization of each band can also be interpreted as the sum of the coupling weights of the mode to  an external field with IP or OP polarization.  Due to the prohibition of interaction between moments with orthogonal orientation $(m\neq m^\prime)$,   IP (OP) modes will couple only to external fields with  IP (OP) symmetry. 
 \begin{figure*}
	\includegraphics[width= .7\textwidth]{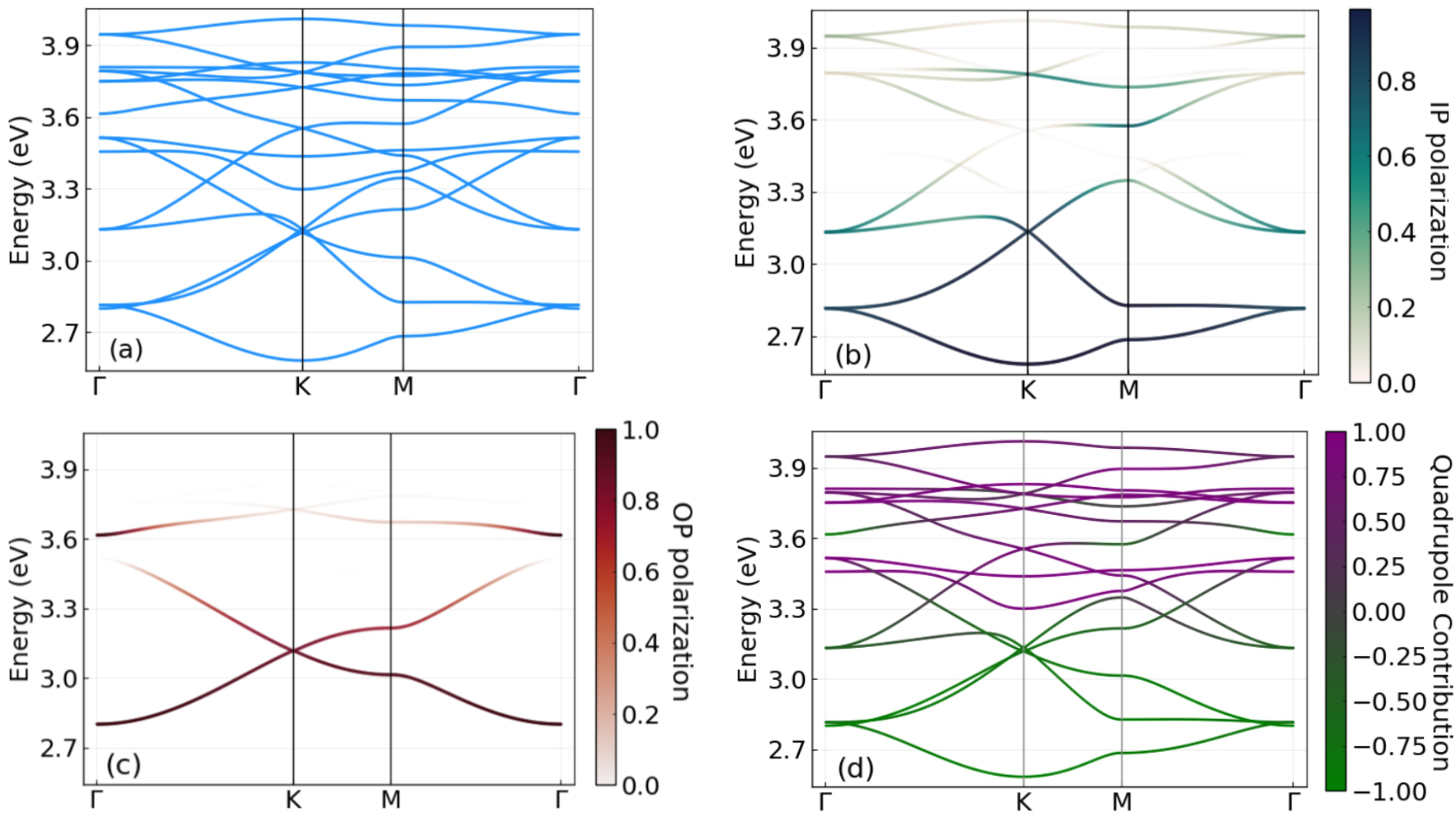}
	\caption{(a) Dispersion relation at the quadrupole level $(l_{max} = 2)$  along the $\Gamma -$~K$ -$~M path of a honeycomb monolayer of $a = 10$ nm Ag nanospheres, using Eq.~\ref{eq:ResFrequDrude}. (b) In-plane  $P_{IP}(s)$  and (c) out-of-plane  polarization $P_{OP}(s)$ component of each state calculated using Eq.~\eqref{eq:polarization0}. (d) Dipole-quadrupole contribution, $ \left< \mathbf{H}^{(quad)}-\mathbf{H}^{(dip)} \right>_s/n_s$, at each state: larger positive values (purple) correspond to larger quadrupole contribution for a given mode, $\ket{s(\vec{k})}$, while larger negative values (green) correspond to larger dipole contribution.    }
	\label{fig:multipolebands}
\end{figure*}

Next, we study the role of multipolar interactions in the honeycomb monolayer. For this case we consider a system of Ag NPs in a host with a larger dielectric constant to lower band energy $\sqrt{\epsilon_h} = 1.9$ with separation between nearest neighbors of $ d = 2.5 a$ so that convergence of the system's eigenmodes is achieved by taking $l_{max} = 2$. At this level, dipole-dipole, dipole-quadrupole, and quadrupole-quadrupole interactions are taken into account. The complete band structure for this system is shown in Fig.~\ref{fig:multipolebands}(a). Due to the smaller separation distance, larger interactions between particles modify the Dirac cone, lifting the degeneracy at the apex as well as shifting bands so that the apex is no longer found within the band gap, as was the case for $l_{max} = 1$. 

Furthermore, a larger number of bands are obtained with respect to the dipole case due to the fact that each multipole moment contributes one band per particle in the unit cell. See the dimension of $\mathbf{H}$ in the Appendix. 
Bands at lower energies, approximately between $2.7-3.3$~eV, resemble those obtained within the dipole approximation. Given the dielectric parameters of the particles described by a Drude dielectric function, at this energy range, it is expected that lower energy bands have a larger dipole character while those at higher energies to have a larger quadrupole contribution. In order to quantify this we notice that due to the orthogonality of the eigenstates $\ket{s}$, any given eigenvalue can be written as $n_s = \braket{s|\mathbf{H}|s} = n_s\braket{s|s} $. 

By separating the different contributions to the interaction matrix $\mathbf{H}$ we can write:
\begin{equation}\label{eq:HdivMono}
n_s = \braket{\mathbf{H}}_s =   \braket{ \mathbf{H}^{\text{dip}}}_s +\braket{ \mathbf{H}^{\text{quad}}    }_s,
\end{equation}
 where we have separated  $\mathbf{H}$ into the different multipolar contributions. 
Matrix $\mathbf{H}^{\text{dip}}$ describes dipole-dipole interactions, while $\mathbf{H}^{\text{quad}}$ describes dipole-quadrupole and quadrupole-quadrupole ones (see the Appendix). It means that by taking only $ \mathbf{H}^{\text{dip}}$, we recover the dipole approximation.  By defining   $ \braket{ \mathbf{H}^{\text{quad}} - \mathbf{H}^{\text{dip}}    }_s/n_S $, we can quantify the dipolar and quadrupolar contribution to each mode, $\ket{s}$, along with its eigenvalue $n_s$, where values close to 1 (-1) mean that the mode is predominantly quadrupolar (dipolar). A value of zero describes a  mode that receives an equal contribution from dipolar and quadrupolar interactions. 

Notice that to calculate the quantities $ \braket{ \mathbf{H}^{\text{dip}}}_s$ and $\braket{ \mathbf{H}^{\text{quad}}    }_s$ we must first solve for eigenvectors $\ket{s}$ of the complete matrix $\mathbf{H}$, we then use the eigenvectors to calculate the contributions coming from each type of multipolar of interaction. These values are calculated for each band along the $\Gamma-K-M$ path and shown in Fig.~\ref{fig:multipolebands}(b). We first corroborate that bands at lower energy are predominately dipolar while those at the highest energies are predominately quadrupolar. Starting from energies at approximately 3.3~eV, bands begin to have a quadrupolar contribution. Surprisingly a large dipolar contribution is found for states at around 3.6~eV near the $\Gamma$-point. We can see that in general bands are neither purely dipolar nor quadrupolar, but the same band can continuously change from being dipolar to quadrupolar.  Due to the bands energy range, we also stress that these results do not take into account interband transitions found in real materials. However, this does not pose a limitation since the optical band structure can be redshifted in energy while maintaining its multipolar character, which is only due to the lattice geometry.

 \begin{figure*} 	
	\includegraphics[width= 1.00\textwidth]{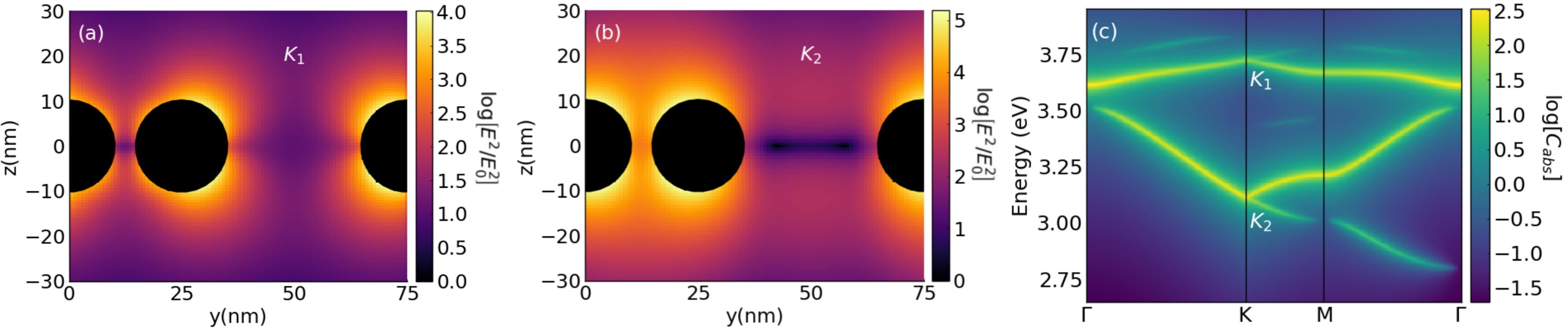} 
	\caption{ Side view of the near electric field intensity enhancement at (a) $K_1$ and (b) $K_2$ taken along the $y$ axis as shown in Fig.~\ref{fig:1},  and  (c) absorption spectrum for out-of-plane illumination.    }
	\label{fig:multipolar3}
\end{figure*}

When multipolar interactions are considered, the restriction prohibiting interaction between modes with different polarization is lifted (see the Appendix). Bands are no longer expected to be either fully IP or OP. Calculation of each band polarization is also shown in Fig.~\ref{fig:multipolebands}(c) and (d). Notice that some bands which are found in the complete band structure in Fig.~\ref{fig:multipolebands}(a),  do not appear in the plots of the  polarization Fig.~\ref{fig:multipolebands}(b)-(c). Since not all modes have a projection onto the dipolar IP or OP base and will therefore have a polarization equal to zero on this basis, in other words, some bands do not couple directly to the dipolar part of the external field but instead require a quadrupolar external field to be directly excited \cite{doi:10.1021/acsphotonics.7b01426}.  We can see that bands are neither fully IP or OP, rather than in general, a band has IP and OP regions.  Mixing of polarization appears predominately at higher energies, which is consistent with the fact that polarization mixing is due to quadrupole interaction.   Notice that there are crossings of  quadrupolar bands at the $K$-point at energies above 3.3~eV that  have polarization different from zero. Despite having a large quadrupolar contribution, they do not require a quadrupolar external field to be excited.

To corroborate the above, we show the absorption spectrum in Fig.~\ref{fig:multipolar3} for OP external illumination. Two things must first say about the excitation of the modes. First, notice that the majority of modes within the Brillouin zone lie outside the light cone and therefore can only be probed experimentally by artificially increasing the momentum parallel to the surface\cite{8852732}. This is typically accomplished using a coupling prism in the attenuated total reflection geometry \cite{Bendana:13} or via near-field probing \cite{Bakker:07} or excitation \cite{doi:10.1021/nn900922z}. Therefore, modes outside the light cone must consider as~eVanescent fields. Secondly, it must clarify that due to the transversality condition, it is not possible to excite the system with a OP polarized electromagnetic wave precisely at the $\Gamma$-point, i.e., at normal incidence. 

In general, extinction lines follow the bands with large OP polarization; however, not all bands are excited with the same intensity. It is because absorption depends not only on the coupling to the external excitation but also on the dielectric properties of the NPs. As a general rule, at smaller values of the imaginary part of the spectral function, there are larger $C_{abs}$, as can be seen from Eqs.~\eqref{green0} and \eqref{eq:Sol1}. It can be seen that the lower energy OP bands in the $\Gamma-K$ region do not appear since dipoles line up parallel to each other and the incident field but perpendicular to the array. The fields they generate, which act upon the neighboring particle, are opposite to the incident field. Therefore they tend to cancel out the neighboring dipole. This effect is small in the $K-M$ and $M-\Gamma$ region because $k_y \neq 0$. Therefore, dipoles are subject to slightly different external field conditions, and they do not completely cancel each other out. To see the difference between modes with strong dipolar and quadrupolar contribution, we calculate the near-field at points $K_2$ and $K_1$, as shown in Fig.~\ref{fig:multipolar3}. We notice that the modes at $K_2$ are almost entirely dipolar, while those at $K_1$ are predominately quadrupolar. At $K_2$, the near-field has an~eVident dipolar symmetry, while at $K_1$, four lobes can be seen to form near the particles' surface, following the quadrupolar symmetry.

\subsection*{Vertically stacked honeycomb lattices}
\begin{figure} 		
 	\includegraphics[width= 0.33\textwidth]{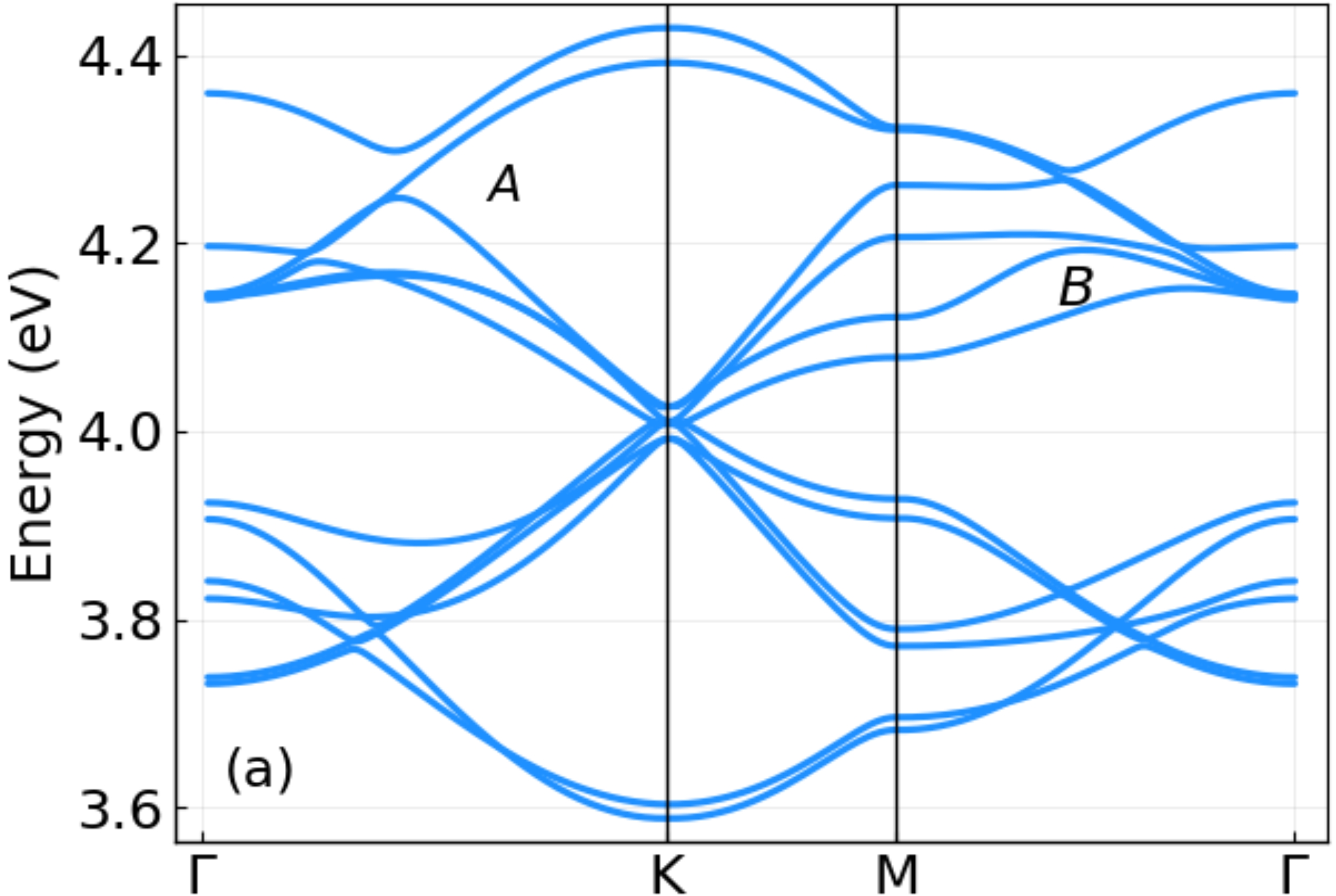}
 	\includegraphics[width= 0.33\textwidth]{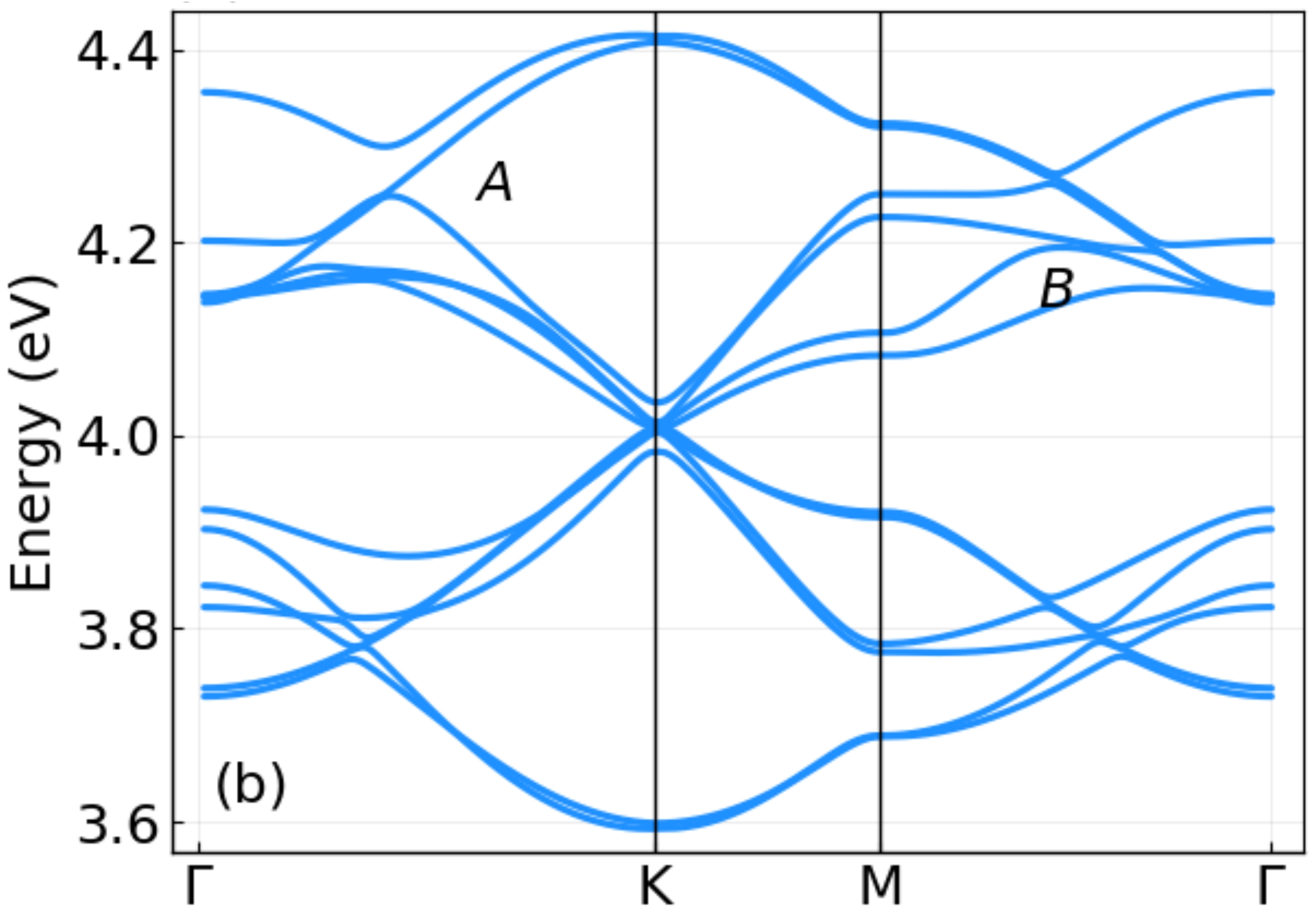}										
	\caption{Dispersion relation for the (a) A-A  and (b) A-B vertically stacked honeycomb plasmonic lattices with separation distance equal to $L = 6a $.}
	\label{fig:BilayerDispersion}
\end{figure}

We next consider the case of a bilayer of honeycomb NPs composed of two vertically stacked monolayers. In contrast to stacked graphene, which has an established separation distance between layers, we are free to vary the separation distance, thus modulating the interaction between layers. To emulate the weak interaction felt between graphene layers, we consider a separation between plasmonic layers to be at least $L \geq 3a$, so that interlayer interaction is fully dipolar and weaker than intralayer interactions. The dispersion relation for a bilayer with A-A and A-B stacking separated by $L = 6a$, in the dipole approximation $l_{max} = 1$, is shown in Fig.~\ref{fig:BilayerDispersion}. As expected, there is double the number of bands concerning the $l_{max} = 1$ monolayer, which is because the unit cell now has twice as many particles. It can be seen that the bilayer band structure resembles the monolayer one and that for both systems the Dirac cone at the $K$ point is preserved. Some interesting differences between the bilayer and monolayer systems are identified. Different from the monolayer, band interaction is now allowed at the crossing points. There are intriguing points at which bands cross each other without interacting, as for the monolayer, and others where bands interact and repel each other. The clearest repulsion points for both types of stacking have been labeled with A and B in Fig.~\ref{fig:BilayerDispersion}. In point A, a clear local gap forms, while in B, the splitting is less visible. Another interesting effect due to vertical stacking is band repulsion at the $K$ point, which causes some of the bands to no longer touch at the apex of the cones. This effect is more noticeable for the A-B than in the A-A stacking. Finally, we note that other than the slightly larger interaction at the $K$ point, the band structure and optical properties between AA and AB type bilayers are similar in and for simplicity, we will work with the AA system. 

\begin{figure}
	\includegraphics[width= 0.36\textwidth]{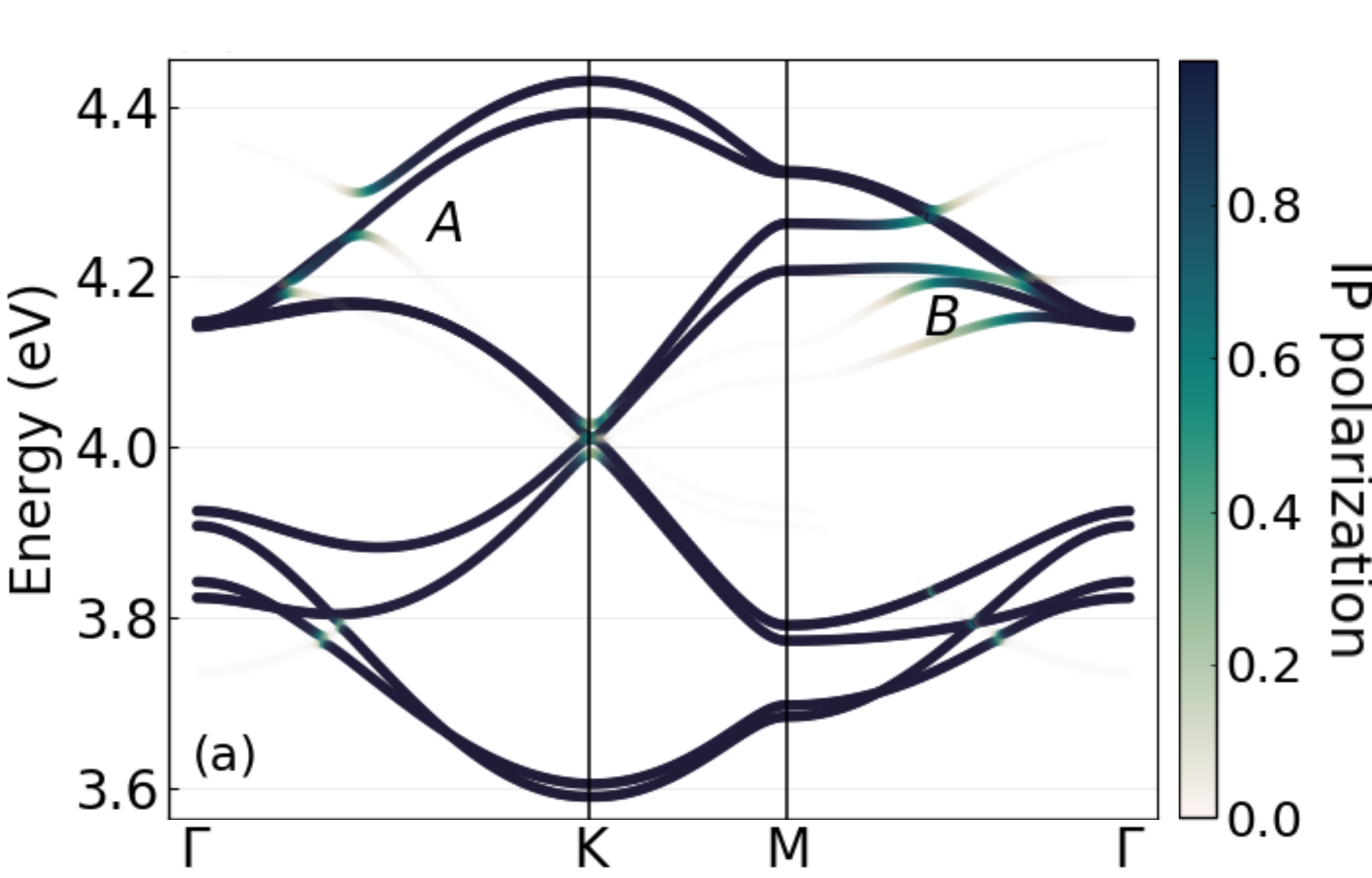}
	\includegraphics[width= 0.36\textwidth]{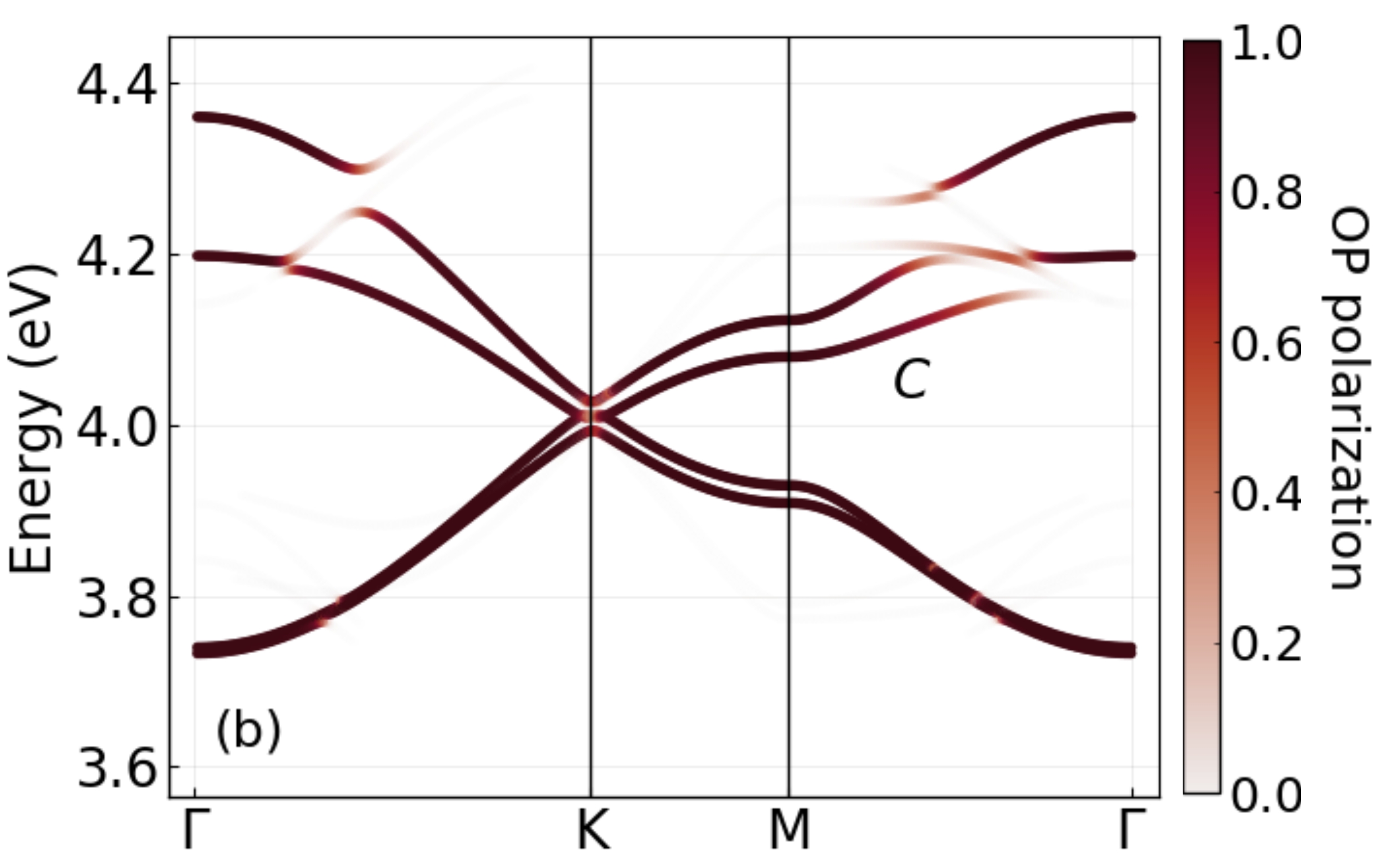}							
	\caption{(a) In-plane  and (b) out-of-plane  polarization of the modes of the A-A stacked bilayer calculated using Eq.~\eqref{eq:polarization0}.  }
	\label{fig:4}
\end{figure}

\begin{figure*}
	\includegraphics[width= 1.00\textwidth]{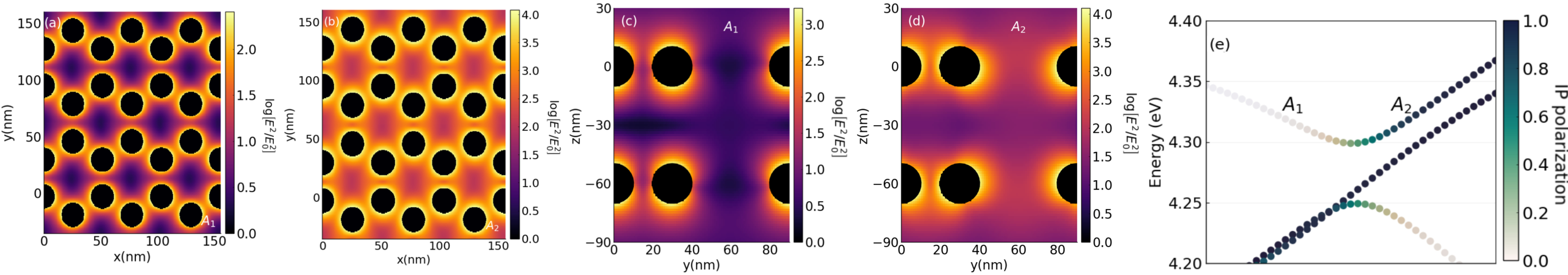}
	\caption{  (a)-(b) Horizontal and (c)-(d) vertical cuts of  near-field enhancement with an in-plane excitation calculated at points labeled as $A_1$ and $A_2$ in (e) near the zone where band repulsion occurs.}
	\label{fig:7}
\end{figure*}

Similar to what was done for the monolayer, we can investigate the IP and OP polarization components along the selected high-symmetry path. It is shown for the AA stacking in Fig.~\ref{fig:4}. Throughout the path, eight bands are predominantly IP while four are predominately OP. As opposed to the monolayer at the dipole level ($l_{max} = 1$), where each band is consistently IP or OP, some bands exhibit an~eVolution of their polarization state along the high symmetry path. As a general rule, polarization varies continuously as the Bloch wave vector sweeps the first Brillouin zone. The vertical anisotropy introduced by the stacking allows interaction between the IP and OP modes belonging to each monolayer, similarly to what is found in phononic crystals \cite{Achaoui_2010}. It is visible both for anticrossing bands and~eVen for the isolated band labeled as C. For the important case of repelling bands, Fig.~\ref{fig:4} shows that the strongly coupled pair of bands (see point A in Fig.~\ref{fig:BilayerDispersion} is composed of a mostly IP polarized band and a mostly OP polarized band and that, away from the anticrossing point, they exchange their polarization state. Fig.~\ref{fig:7} illustrates in more detail how the polarization is transferred between the two repelling bands at point A.

\begin{figure}
	\includegraphics[width= 0.35\textwidth]{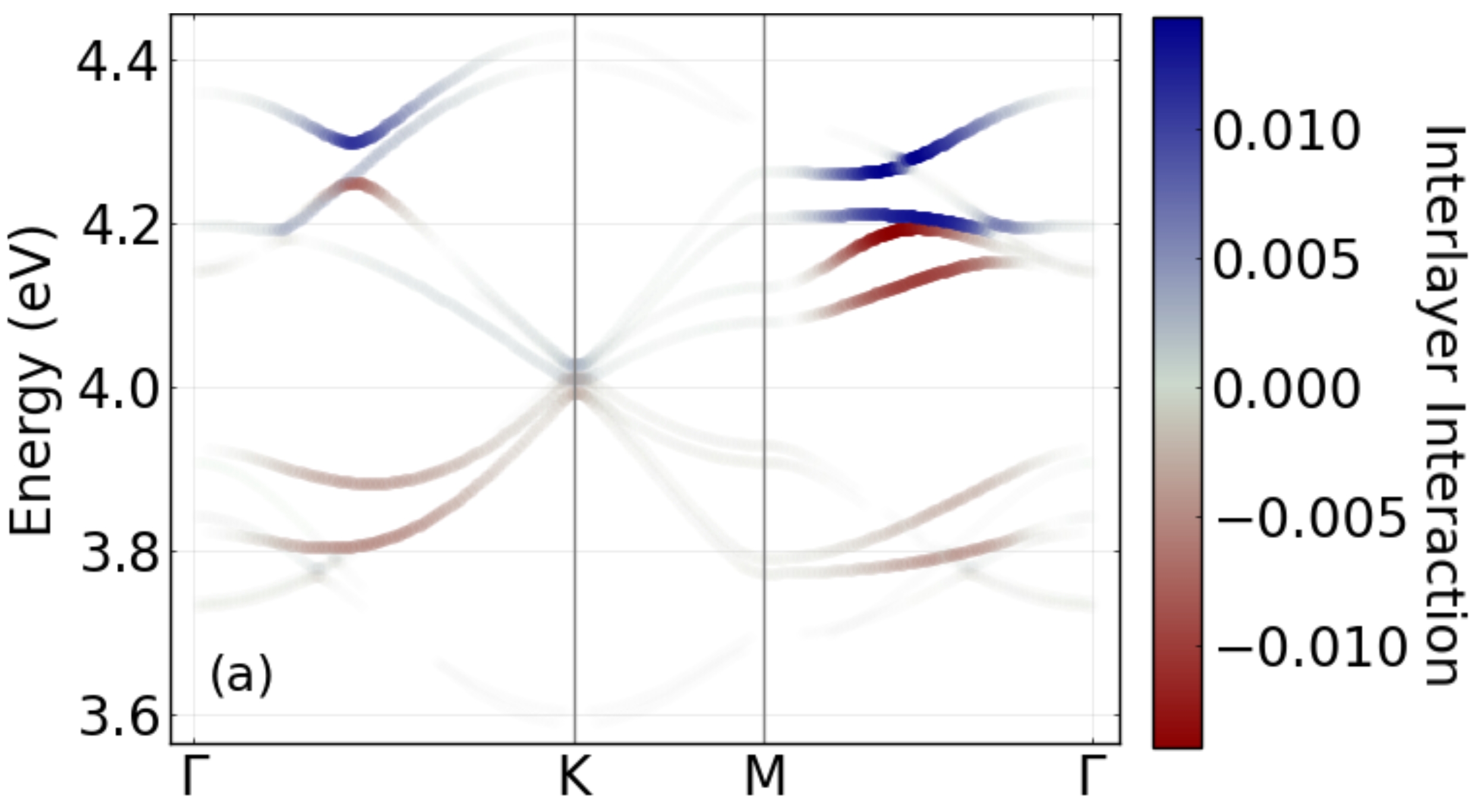}
	\includegraphics[width= 0.35\textwidth]{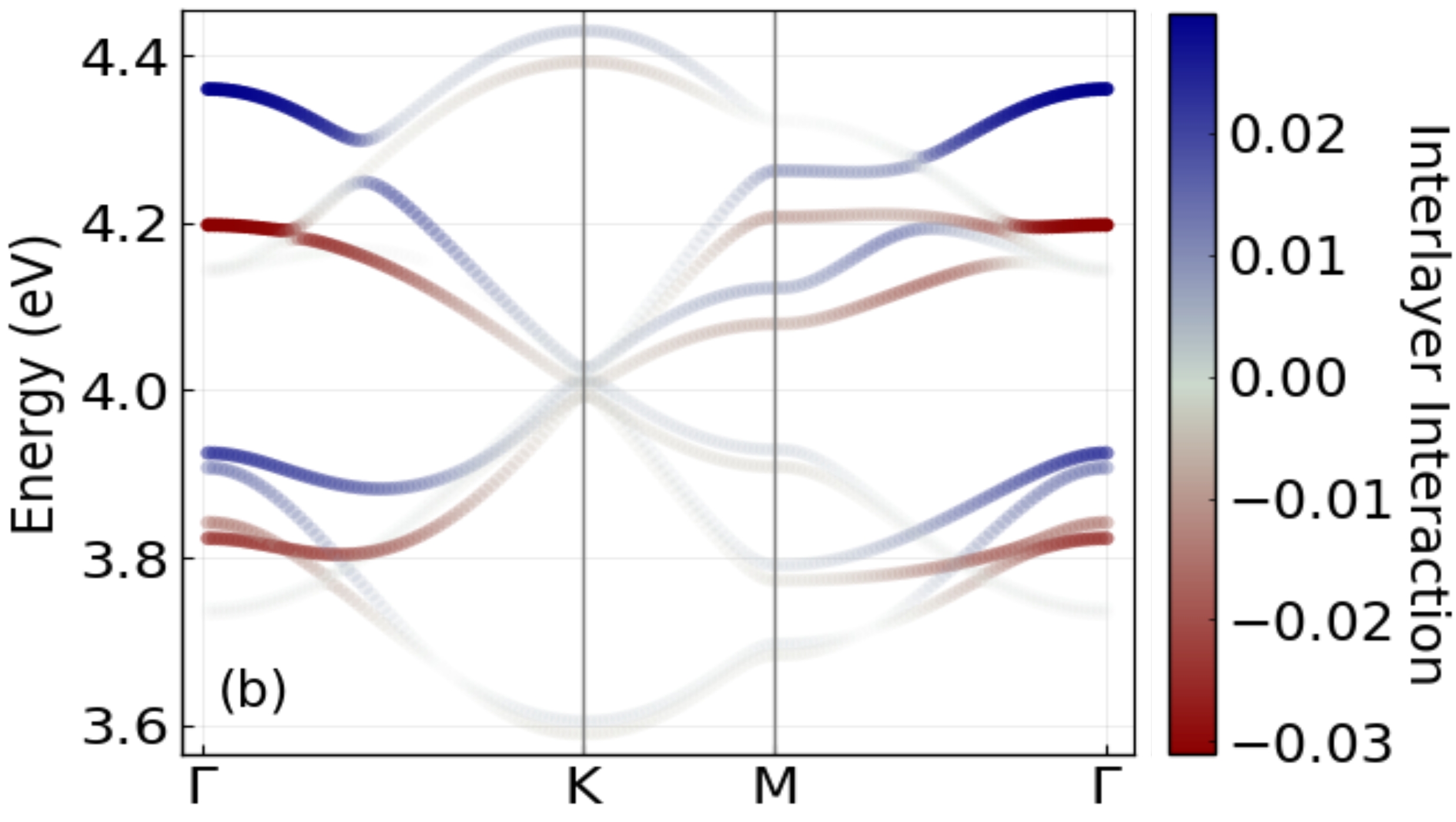}    
	\caption{ Decomposition of the interaction matrix with relation to $n_s$  along high symmetry trajectories given by $\left<\mathbf{H}_{i}^{(inter)} \right>_s/n_s$ with $i = 1,2$. Contribution from (a) IP-OP $H^{\text{(inter)}}_2$ and (b) IP-IP/OP-OP  Interlayer interactions $H^{\text{(inter)}}_1$. }
	\label{fig:5}
\end{figure}

To further investigate the properties of the band structure of the vertical stack and the role of inter- and intralayer interactions we write the system eigenvalues as
\begin{equation}\label{eq:Hdiv}
n_s = \braket{\mathbf{H}^{(\text{vs})}}_s = \mathbf{n}_{1,0}+\braket{ \mathbf{H}^{(\text{inter})}}_s +\braket{ \mathbf{H}^{(\text{intra})}    }_s
\end{equation}
where $\mathbf{H}^{(\text{vs})}$, is the interaction matrix of the vertically stacked system, $\mathbf{n}_{1,0}$ is a diagonal matrix with components of the isolated dipole eigenvalue $n_{1,0} = 1/3$, where we separate the interaction between particles belonging to the same layer, $\mathbf{H}^{\text{(intra)}}$, from the one between particles of different layers, $\mathbf{H}^{\text{(inter)}}$. The interlayer interaction can be written as  $\mathbf{H}^{\text{(inter)}} = \mathbf{H}_{1}^{\text{(inter)}} + \mathbf{H}_{2}^{\text{(inter)}}$,   where $\mathbf{H}_{1}^{\text{(inter)}} $ describes interactions between pairs of particles having charge distributions with either IP-IP or OP-OP symmetry  and  $\mathbf{H}_{2}^{\text{(inter)}} $ describes interaction between charge distributions with IP-OP symmetry  (see the Appendix). Therefore the quantity $\braket{ \mathbf{H}_1^{(\text{inter})}    }_s$ is a measure of the contribution of IP-IP/OP-OP interlayer interaction to mode $\ket{s}$. Likewise, $\braket{\mathbf{H}_{2}^{\text{(inter)}}}_s$ measures the contribution from IP-OP interlayer interaction. These values are calculated for each band along the $\Gamma-K-M$ path in Fig~\ref{fig:5}. Interestingly, interactions can change sign for different energy values at a given $k$ point. This is particularly noticeable for IP-OP interlayer interaction in the vicinity of IP-OP anticrossings where band repulsion takes place (see regions A and B in Fig.~\ref{fig:BilayerDispersion}). Although this type of interlayer interaction is in general small, in fact contributing only 1\% to $n_s$, it is enough to cause band repulsion. This can be explained by noticing that at the crossing points the intralayer interaction, which in general contributes around 10\% to $n_s$, goes almost to 0 (see the Appendix). It can be concluded that band repulsion and the formation of mini-gaps at the K-point is a result of IP-OP interlayer interaction due to the anisotropy caused by vertical stacking.

We next investigate the properties of the absorption spectrum and the near electric field of the eigenmodes of the vertical stack. 
\begin{figure}
	\includegraphics[width= 0.35\textwidth]{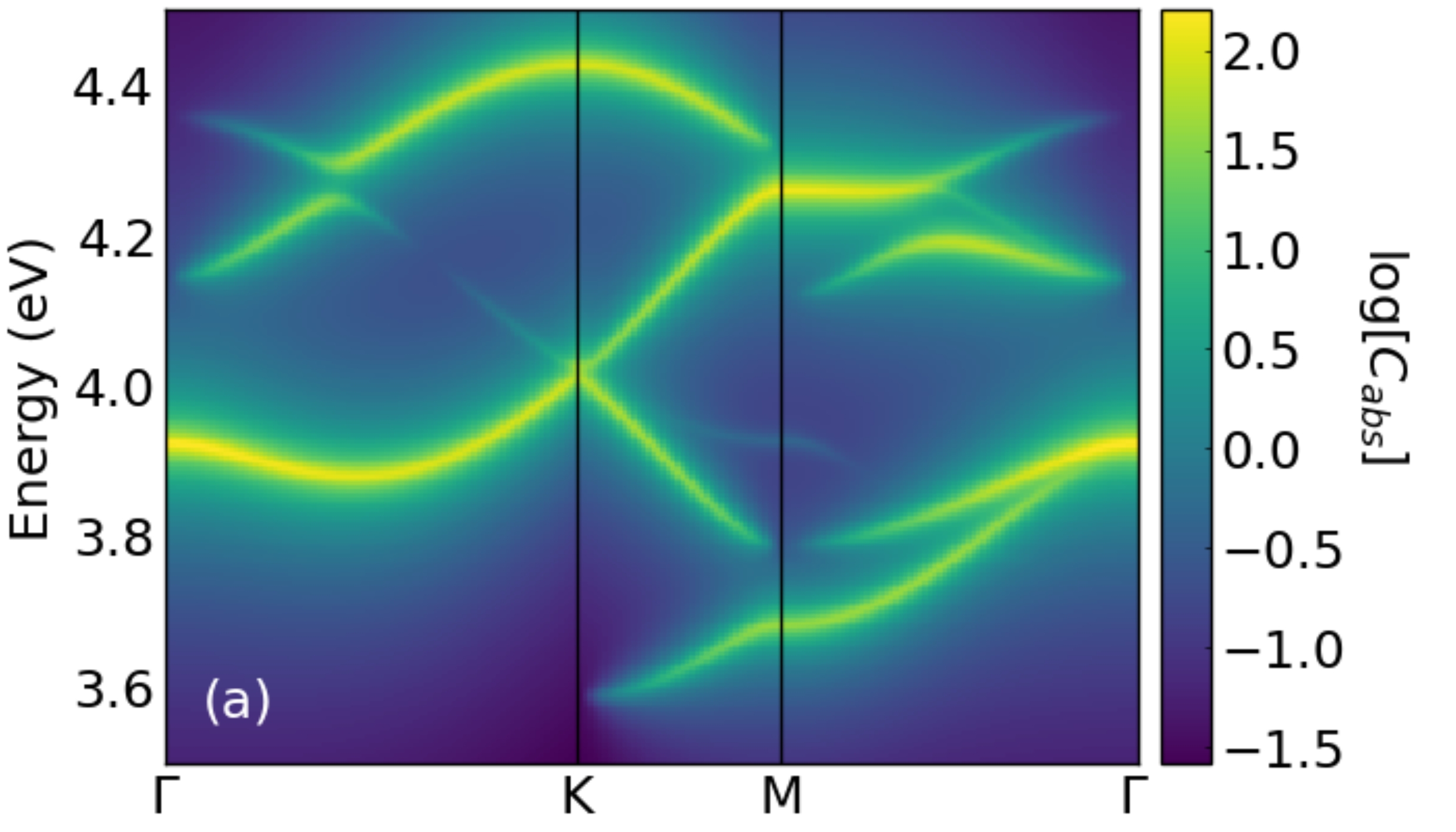}
	\includegraphics[width= 0.35\textwidth]{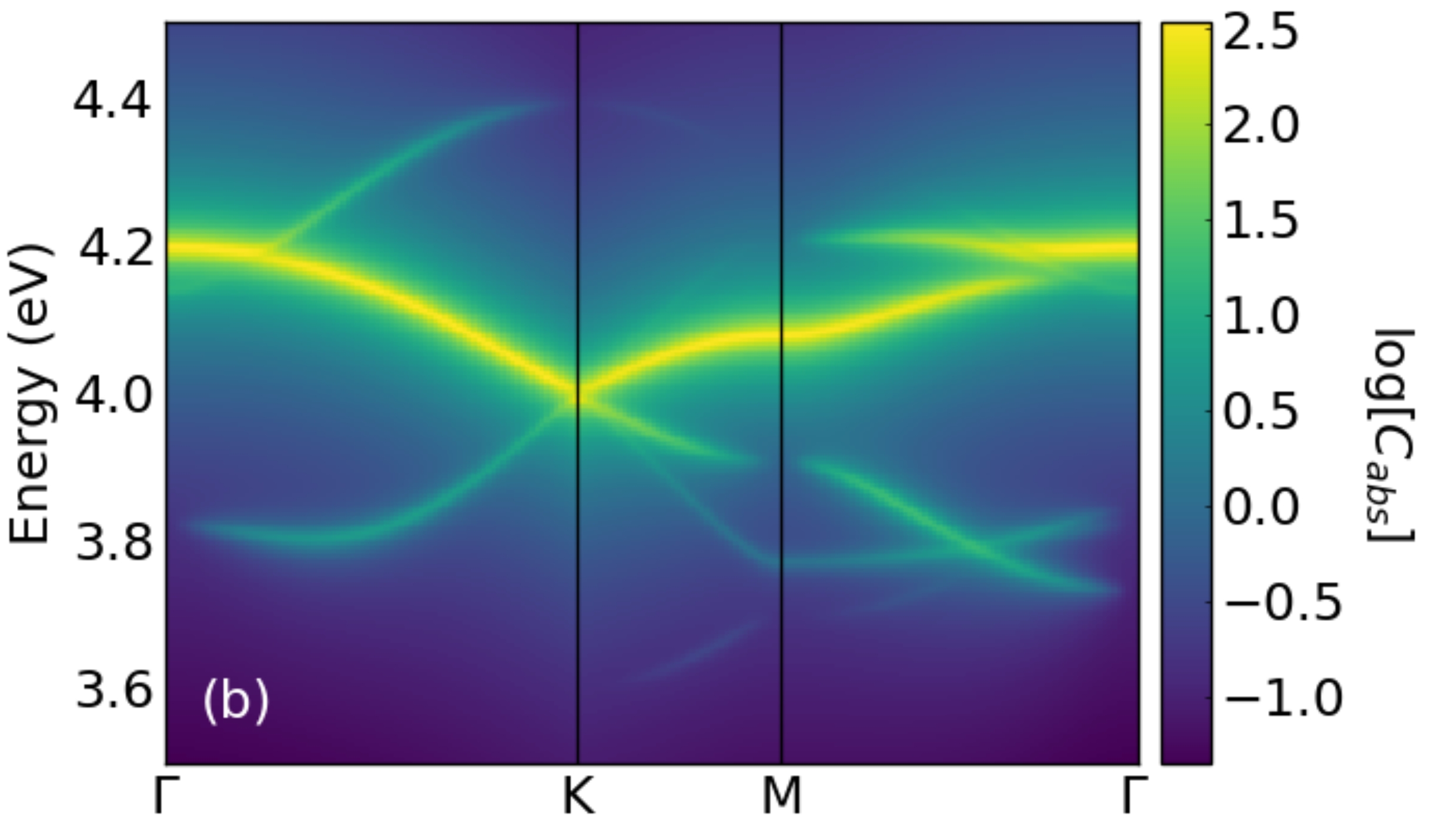}
	\caption{Near field intensity spectrum in $\log_{10}$ scale for (a) in-plane and (b) out-of-plane excitation of a vertical AA stacking of honeycomb lattices at the dipole level.}
	\label{fig:bilaterAbs}
\end{figure}

In Fig.~\ref{fig:bilaterAbs}, it can be seen that the absorption spectrum follows the band structure of the system, similar to the monolayer. Remarkably, only some of the bands appear as bright in the absorption spectrum. Like the monolayer case, this optical band selection is a consequence of the selection rules dictated by the matrix A symmetry (see the Appendix). We stress that, due to the IP-OP coupling between layers, it is possible to excite OP modes with IP illumination and vice versa (compare with Fig.~\ref{fig:4}). The band repulsion takes place in points A and B, as they are visible in the absorption spectrum, while the width of the bands hides the mini-gaps. We checked that by fictitiously switching off IP-OP interlayer interaction, $\mathbf{H}_{2}^{\text{(inter)}} = 0$, the excitation of modes with different symmetry to that of the exciting field is not allowed (not shown here). It is similar to the case of two independent monolayers. Recall that losses are considered in the Drude model (scattering rates) and so, included in the real and imaginary values of the spectral variable of Eq.~\ref{eq:res_condition}. Thus, bands obtained from Eq.~\ref{eq:ResFrequDrude} also include losses. The influence of losses is visible in the near field spectra. When the scattering rates increase, band thickness becomes more diffuse, and the maximum of the near field intensity diminishes. From the theoretical point of view, bands are precisely determined to always make visible mini gaps and deviations. Experimentally, bands can spread out over an area, making those band deviations hard to recognize unless compared to a theoretical model. 

Finally, to visualize the polarization switch, we plot the near electric field within the vertical stack in the vicinity of a band repulsion and the $K$ point. In Fig.~\ref{fig:7}, we show a close up of the region near band repulsion labeled A (see Fig.~\ref{fig:BilayerDispersion}), as well as the near-field enhancement due to IP external excitation at two different points along with the band. In $A1$ ($A2$), the band has large OP (IP) polarization. It can be seen that the near-field enhancement is nearly one order of magnitude larger when the exciting field and the band have matching polarizations. However, due to the interlayer coupling, enhancement is not negligible at points with large OP coupling such as A1. Furthermore, a clear rotation in the dipole orientation associated with each particle is observed while following the band. At point A1, the polarization is more OP oriented despite being excited by an IP field. On the other hand, at point $A2$, the near-field is more IP following the type of excitation.  

\section{Conclusions}
The honeycomb lattice is the simplest and most attractive bipartite, non-Bravais lattice, whose properties are investigated across several scales. Ordered resonant nanoparticle arrays constitute a suitable platform for investigating similarities and differences between the electronic band structure of the atomic scale and the optical band structure of nanoscale honeycomb lattices. Here, we have studied the modification of the optical band structure, optical absorption, and spatial near-field distribution of a honeycomb plasmonic lattice introduced by stacking two such lattices at subwavelength distances. We used the multipolar spectral representation (MSR) method to clarify similarities and differences between the optical and atomic potential within a tight-binding type of model. In the band structure of a plasmonic monolayer, we highlight Dirac cones formation at the K-point due to in-plane modes, besides the expected ones related to the out-of-plane modes, akin to the out-of-plane p-bands in graphene. This remarkable difference with what is found in graphene stems from the limitation of the analogy between atomic orbitals and multipole moments of the single plasmonic nanosphere. The fact that particles interact predominately through their dipole moment, which has a symmetry similar to p-orbitals, does not allow for the sp2 hybridization typical of the in-plane carbon bonds. 

The MSR method permits the precise description of the interlayer and intralayer coupling among in-plane and out-of-plane polarized modes in the stacked system. We have shown that the material anisotropy introduced by the layers vertical stacking introduces the coupling of all polarization components. One of the main results we found is local gaps resulting from bands avoided crossing with opposite symmetry and belonging to different layers. The strong mode coupling manifests itself as in-plane -- out-of-plane polarization mixing, causing the near-field intensity spatial redistribution. By leveraging the in-plane field component, the enhancement and localization of the electromagnetic field within the vertical stack can be increased, which may be useful in the context of open cavities and strong light-matter interaction.

\begin{acknowledgements}
We acknowledge partial support from CONACyT projects 1564464 and 1098652; and from DGAPA-UNAM projects PAPIIT IN107319 and IN109618.
\end{acknowledgements}

\appendix

\section{Multipolar Spectral Representation}
The periodic lattice can be spanned by translation of a unit cell  through  vectors $\vec{t}_1$ and $\vec{t}_2$, such that the coordinates of each particle in the lattice are
	\begin{equation}
	\vec{R}_{M,N,i} = M\vec{t}_1 + N\vec{t}_2 + \vec{b}_i 
	\end{equation}
	where $M,N$ label the cell in which the particle is found and $\vec{b}_i$ gives the location of the i-th particle in the unit cell. Therefore each particle in the lattice is defined by the set of indexes $(M,N,i)$ with $i = 1,...,N_B$, where $N_B$ is the number of particles per unit cell. For example, in the case of a honeycomb monolayer $N_B = 2$.
	If the NPs are sufficiently close to each other their near-field couple due to Coulomb interactions between the induced charges on each particle, which are described with a multipolar expansion. To completely identify the multipole induced on a given particle we use the set of indexes $(l,m,M,N,i)$, where $lm$ is the multipole order used to describe the induced charge and $(M,N,i)$ identify the particle in the lattice. To simplify notation, throughout this work all variables with multipolar dependence are represented as vectors, where each component of the vector is identified by the set $(l,m,M,N,i)$ which  is also represented by $\mu$. Using this notation, the  multipolar moments ${Q}_{\mu}$ induced on a sphere in the presence of a frequency-dependent potential can be expressed as

\begin{equation}\label{multmom}
\begin{split}
{Q}_{\mu}&=-{\alpha}_{\mu}  {V}_{\text{tot}}^{\mu}\\
&=-{\alpha}_{\mu} \left[{V}_{\text{ext}}+ {V}_{\text{ind}}\right]^{\mu},
\end{split}
\end{equation}
where ${\alpha}_{\mu}(\omega)$ are the frequency-dependent multipolarizabilities. The total potential $V_{\text{tot}}$ is divided into  $ {V}_{\text{ext}}^{\mu} $    and    ${V}_{\text{ind}}^{\mu}$,  the external and induced potential felt by the nanospheres, respectively. 

Due to the periodicity of the lattice, we seek Bloch-like solutions of Eq.~\eqref{multmom} for the induced multipole moments of each particle: 
\begin{equation}\label{eq:BlochSol}
\begin{split}
Q_{lm,MNi}&= Q_{lm,i00} \text{exp}[-j \vec{k}\cdot \vec{R}_{MNi} ], \\
\end{split}
\end{equation}
where $\vec{k}$ is the wavevector of the external potential. Multipolar moments $Q_{lm,MNi}$ in cell $M,N$ can be seen as replicas of moments $Q_{lm,00i}$ phase shifted by a term $\text{exp}[-i \vec{k}\cdot \vec{R}_{MNi} ]$. By using Eq.~\eqref{eq:BlochSol} in \eqref{multmom} and applying periodic conditions  we can  solve for the multipolar moments within only one unit cell 
%\begin{widetext}
\begin{equation}\label{eq:TBSR1}
Q_{lm,i}  =- \alpha_{lm,i} \left[V_{lm,i}^{\text{ext}} + \sum_{i^\prime}^{\substack{\text{nearest} \\ \text{neighb.} }}\sum_{l^\prime m^\prime }  {A}_{lm i}^{l^\prime m^\prime i^\prime} {Q}_{l^\prime m^\prime, i^\prime} e^{j\vec{k}\cdot\vec{d_{ii^{\prime}}}} \right] 
\end{equation}
%\end{widetext}
where $d_{ii^{\prime}} = (\vec{R}_{00i} -\vec{R}_{M^\prime N^\prime i^\prime} )$ with adequate $M,N$ indexes. 

For the case of an array of nanospheres it is useful to use spherical coordinates and expand the potentials and components of the interaction matrixes in spherical harmonics\cite{PhysRevB.22.4950}.  In this basis the multipolarizability  is given as \cite{PhysRevB.34.3730}:
\begin{equation}\label{polariz}
\alpha_{\mu}(\omega) = \frac{2l+1}{4\pi} \frac{n_{0l}}{n_{0l}-u (\omega)}a_{MNi}^{2l+1},
\end{equation}
where $n_{0l} = l/(2l+1)$. Notice that $n_{0l}$ represents the eigenvalues of the isolated multipole of order $l$, and does not depend on  $m$ due to the symmetry of the particle.

\begin{widetext}
\section{Interaction matrix in the nearest-neighbor approximation}
The $\mu$-th term of the potential felt by particle $i$ due to the induced moments on particle $i^\prime$ as
\begin{equation}
{V}_{\text{ind}}^{\mu}  = \sum_{\mu^\prime}{A}_{\mu}^{\mu^\prime} {Q}_{\mu^\prime},
\end{equation}
Where elements of the coupling matrix $\mathbf{A}$ are written as:
\begin{equation}
\begin{split}
\bm{\mathrm{{A}}}^{l^{\prime} m^{\prime} j}_{lmi}   &= (-1)^{l^{\prime}+m^{\prime}} \frac{\left[Y(\theta_{ij},\phi_{ij})_{l+l^{\prime}}^{m-m^{\prime}} \right] ^{\ast} }{ R_{ij}^{l+l^{\prime}+1} }\times \\
&\times \left[ \frac{(4\pi)^{3}}{(2l+1)(2l^{\prime}+1)(2l+2l^{\prime}+1)} \frac{(l+l^{\prime}+m-m^{\prime})! (l+l^{\prime}-m+m^{\prime})!}{(l^{\prime}+m^{\prime})!(l+m)!(l^{\prime}-m^{\prime})!(l-m)!} \right] ^{1/2} \label{Amat}
\end{split} 
\end{equation}
\end{widetext}
where $R_{ij}= | \vec{R}_{j}-\vec{R}_{i}| $ is the vector that joins the centers of particles $i$ and $j$ and , $i\neq j$, and $Y_{l+l^{\prime}}^{m-m^{\prime}} (\theta_{ij},\phi_{ij})  $  are spherical harmonics~eValuated at angles $\theta_{ij},\phi_{ij},$ corresponding to $R_{ij}$. Therefore, the elements $A_{\mu}^{\mu^\prime}$ describes the potential generated by the $lm$-th multipole at the location described by indexes $M,N,i$ and felt by $l^\prime m^\prime$ multipole moment on particle located at $M^\prime,N^\prime,i^\prime$.

\begin{figure}[h]
	\includegraphics[width= 0.32\textwidth]{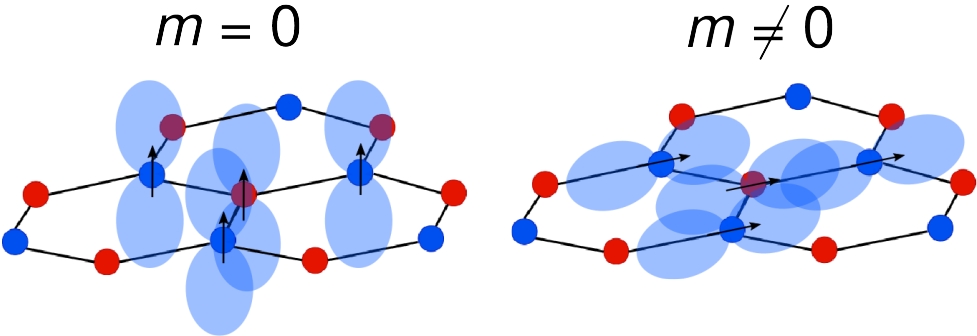}
	\caption{Schematic of  out-of-plane $(m \neq 0)$ and in-plane (m=0) modes of a monolayer in the dipole approximation $l_{max} = 1$. Modes have no interaction between orthogonally oriented moments.  }
	\label{fig:app1}
\end{figure}

It is important to note that for a system of particles on a plane the spherical harmonic in  Eq.~\eqref{Amat} satisfies
\begin{widetext}
\begin{equation}
\begin{split}
Y_{l+l^{\prime}}^{m-m^{\prime}}( \pi/2,\phi_{ij}) =& \frac{(l+l^\prime + m - m^\prime )\mod 2}{2}(-1)^{\frac{l+l^\prime + m - m^\prime }{2}}e^{j(m - m^\prime)\phi_{ij} }\\ 
&\times\sqrt{\frac{(2l+2l^\prime+1)(l+l^\prime + m - m^\prime-1 )! (l+l^\prime - m + m^\prime -1)!}{\pi(l+l^\prime + m - m^\prime )!! (l+l^\prime - m + m^\prime )!!}}
\end{split}
\end{equation}
Therefore in the dipole approximation, interaction between moments with different IP $(m = \pm 1)$ and OP $(m = 0)$ symmetry is forbidden as illustrated in the schematic of Fig.~\ref{fig:app1}. 
 We  then assume an external potential of the form 
\begin{equation}\label{eq:ExtPot}
V_{lm,MNi}^{\text{ext}} = V_{lm,i}^{\text{ext}} \text{exp}[j(\omega t - \vec{k}^\prime \cdot \vec{R}_{MNi} )]
\end{equation}
Using Eqs. \eqref{eq:BlochSol} and \eqref{eq:ExtPot}  in Eq. \eqref{multmom} we obtain a relation for the multipolar moments
%\begin{widetext}
\begin{equation}\label{eq:TBMRS1}
\begin{split}
{Q}_{lm,MNi} & =-{\alpha}_{lm,i} \left[{V}_{\text{ext}}+ {V}_{\text{ind}}\right]^{lm,MNi}\\	
Q_{lm,i}e^{-j\vec{k}\cdot \vec{R}_{MNi}} & =- \alpha_{lm,i} \left[V_{lm,i}^{\text{ext}} e^{-j\vec{k}^\prime \cdot \vec{R}_{MNi}}+ \sum_{M^\prime N^\prime i^\prime}\sum_{l^\prime m^\prime }  {A}_{lmn}^{l^\prime m^\prime n^\prime} {Q}_{l^\prime m^\prime, M^\prime N^\prime i^\prime} e^{-i\vec{k}\cdot \vec{R}_{M^\prime N^\prime i^\prime}} \right]  \\
\end{split}
\end{equation}

%\end{widetext}

\begin{equation}\label{eq:TBSR1_app}
Q_{lm,i}  =- \alpha_{lm,i} \left[V_{lm,i}^{\text{ext}} + \sum_{M^\prime N^\prime i^\prime}\sum_{l^\prime m^\prime }  {A}_{lm i}^{l^\prime m^\prime i^\prime} {Q}_{l^\prime m^\prime, i^\prime} e^{j\vec{k}\cdot(\vec{R}_{00i} -\vec{R}_{M^\prime N^\prime i^\prime} )} \right] 
\end{equation}
\end{widetext}

Notice that the sum over indexes $(M^\prime N^\prime i^\prime)$ describes interaction with all particles of the lattices. The tight-binding approximation limits this sum to include only the nearest neighbors. Furthermore, due to periodic conditions, which intuitivly cand be understood as  the assumption of equivalence between particles in different cells, allows us to only have calculate the induced multipole moments in one unit cells, say $M = 0, N = 0 $. so Eq.~\eqref{eq:TBMRS1} can be reduced to
\begin{widetext}
\begin{equation}\label{eq:TBSR2}
Q_{lm,i}  =- \alpha_{lm,i} \left[V_{lm,i}^{\text{ext}} + \sum_{M^\prime N^\prime i^\prime}^{\substack{\text{nearest} \\ \text{neighbors} }}\sum_{l^\prime m^\prime }  {A}_{lmn}^{l^\prime m^\prime n^\prime} {Q}_{l^\prime m^\prime, M^\prime N^\prime i^\prime} e^{j\vec{k}\cdot(\vec{R}_{00i} -\vec{R}_{M^\prime N^\prime i^\prime} )} \right] 
\end{equation}
%\end{widetext}
This equation can then be brought to a matrix form equivalent to Eq.~\eqref{secular} 
\begin{equation}
\left( u(\omega)\mathbf{I} + \mathbf{H}\right) \vec{x} = \vec{F},
\end{equation}
with matrix elements of $\mathbf{H}$
%\begin{widetext}
\begin{equation}\label{eq:Hmat}
	\pmb{H}^{l^\prime m^\prime i^\prime }_{lm i}(\vec{k}) =  n_0^l \delta^{l^\prime m^\prime }_{lm}+ (-1)^{l^\prime}\frac{\sqrt{l l^\prime a^{2l+1}a^{2l^\prime+1} }}{4\pi} \sum_{ i^\prime  }^{nn} A^{l^\prime m^\prime  i^\prime  }_{lm i}e^{j \vec{k}\cdot \vec{R}_{i i^\prime}} 
\end{equation}
\end{widetext}

\section{Coupling weights}
To solve Eq.~\eqref{eq:BlochSol} we first need to diagonalize matrix $\mathbf{H}$. We find the unitary matrix $\mathbf{U}$ that satisfies
\begin{equation}
	\mathbf{U}^\dagger \mathbf{H} \mathbf{U} = \mathbf{n}
\end{equation} 
where $\mathbf{U}$ is a matrix whose columns are the system's eiegenvectors,	$\mathbf{U}^\dagger$ is the transpose and complex conjugate of $\mathbf{U}$ and $\mathbf{n}$ is a diagonal matrix whose elements are the eigenvalues of $\mathbf{H}$. Solutions to  Eq.~\eqref{eq:BlochSol} can the be found with the Green matrix formed by the elements,
\begin{equation}
	G_{\mu}^{\mu^\prime}(\vec{k},\omega) = -\sum_{s} \frac{U_{\mu}^{s}(\vec{k}) \left[U^{\mu^\prime}_s(\vec{k})\right]^{\dagger}}{u(\omega)-n_s}.
\end{equation}
The coupling strength is defined as ther term $C_{\mu}^{\mu^\prime}(s) = U_{\mu}^{s}(\vec{k}) \left[U^{\mu^\prime}_s(\vec{k})\right]^{\dagger}$. Interpretation of the coupling strength can be made more clear for example when calculating the induced multipole moments due to an external field as in Eq.~\eqref{eq:Sol1}. We see that coupling strenght $C_{\mu}^{\mu^\prime}(s)$ describes the contribution to moment $\mu$ form an external field $V^{ext}_{\mu^\prime}$ through mode $s$.

\section{Monolayer Interaction Matrix $l_{max} = 2$}
The matrix elements of $\mathbf{H}$ for a monolayer in the quadrupole approximation are given by Eq.~\eqref{eq:TBSRHelement}. Recall that this maxtrix includes dipolar-dipolar, dipolar-quadrupolar and quadrupolar-quadrupolar interactions. The total matrix will have dimension of $(2\times 8)\times(2\times 8)$ since there are two particles in each cell and since $l_{max} = 2$  there are eight multipole moments that must be described with their corresponding interactions among them. The matrix will have the following form: 
\begin{equation}
\mathbf{H} = \begin{pmatrix}
n_0 &0 \\
0& n_0 
\end{pmatrix}   + \begin{pmatrix}
0 & H(P_1 P_2)\\
H(P_2 P_1)& 0 
\end{pmatrix}
\end{equation}
where $\mathbf{n_0}$ is a $8\times8$  diagonal matrix whose first three components are the eigenvalues of an isolated dipole and the last five are those of an isolated quadrupole.  $ H(P_1 P_2)$ is a  $8\times8$ submatrix that describes interaction between particles $P_1$ and type $P_2$ as shown in the schematic of Fig.~\ref{fig:1}. We can further divide $ H(P_1 P_2)$ by the type of interactions in the following manner 
\begin{equation}
 \mathbf{H}(P_1 P_2) =  \begin{pmatrix}
 H^{dip}_{dip} & H^{quad}_{dip} \\
 H^{dip}_{quad} & H^{quad}_{quad} \\
 \end{pmatrix}
 \end{equation}
where $ H^{dip}_{dip}$ is a $3\times3$, $H^{quad}_{dip}$, is a $3\times 5$ and $H^{quad}_{quad}$, is a $5\times5$, matrix describing dipole-dipole, dipole-quadrupole and quadrupole-quadrupole interactions respectively, between particles 1 and 2 .

\section{Vertical Stack Interaction Matrix $l_{max} = 1$}
The matrix elements of $\mathbf{H}$ are given in Eq.~\eqref{eq:TBSRHelement}. The element $H_{\mu}^{\mu^\prime} = H_{lm,MNi}^{l^\prime m^\prime ,M^\prime N^\prime i^\prime }$, describes the interaction between the $lm$-th multipole of particle $MNi$ with the $l^\prime m^\prime$-the multipole of particle $M^\prime N^\prime i^\prime$. In this sense, the interaction matrix of a honeycomb monolayer system can be divided in the following manner.
\begin{equation}
	\begin{split}
\mathbf{H}^{(mono)} & = \begin{pmatrix}
n_0 &0 \\
0& n_0 
\end{pmatrix}   + \begin{pmatrix}
0 & H(P_1 P_2)\\
 H(P_2 P_1)& 0 
\end{pmatrix} \\
 & = \mathbf{n}_0 + \mathbf{H}^{(intra)}, 
\end{split}
\end{equation}
where $ H(P_1 P_2)$ is a submatrix that describes interaction between particles $P_1$ and type $P_2$, as shown in the schematic of Fig.~\ref{fig:1} and $n_0$ is a diagonal submatrix whose components are the eigenvalues of an isolated dipole. 

\begin{figure}
	\includegraphics[width= 0.32\textwidth]{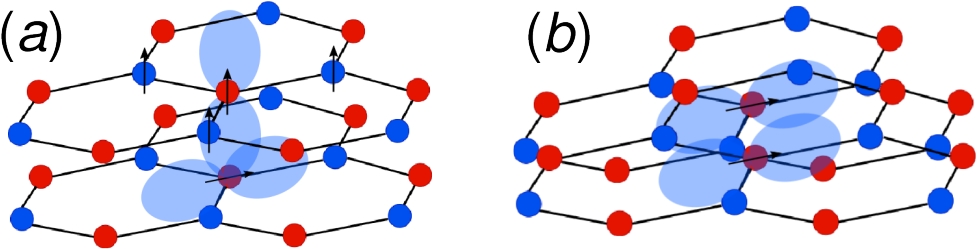}
	\caption{ Schematic of different types interlayer interaction a) $H^{\text{(inter)}}_2$ showing the case of OP-IP  and b) $H^{\text{(inter)}}_1$  a case of IP-IP.   }
	\label{fig:app2}
\end{figure}

For two vertically staked layers the interaction matrix can be  divided as
\begin{widetext}
\begin{equation}
	\begin{split}
\mathbf{H}^{(vs)} & = \begin{pmatrix}
n_0 & H^{(inter)}(P_1 P_2) & H^{(intra)}(P_1 P^\prime_1) & H^{(intra)}(P_1 P^\prime_2) \\
 H^{(inter)}(P_2 P_1) & n_0 &   H^{(intra)}(P_2 P^\prime_1) & H^{(intra)}(P_2 P^\prime_2)  \\
 H^{(intra)}(P^\prime_1 P_1) & H^{(intra)}(P^\prime_1 P_2) & n_0 &  H^{(inter)}(P^\prime_1 P_2) \\
  H^{(intra)}(P^\prime_2 P_1)& H^{(intra)}(P^\prime_2 P_2) & H^{(inter)}(P^\prime_2 P_1) & n_0
\end{pmatrix} \\
& = \mathbf{n}_0 + \mathbf{H}^{(intra)} + \mathbf{H}^{(inter)} 
\end{split}
\end{equation}
\end{widetext}
where, for example, $H^{(intra)}(P_1 P_2)$ is a matrix describing interaction between particles $P_1$ and $P_2$ in the same layer, and $H^{(inter)}(P_1 P^\prime_2)$ describes interaction between particles of different layers. Matrix $\mathbf{H}^{(inter)}$ can further be separated into IP-IP and OP-OP interactions as in $\mathbf{H}_1^{(inter)}$ and IP-OP interactions as in $\mathbf{H}_2^{(inter)}$ of Eq.~\eqref{eq:Hdiv}. Specifically, $\mathbf{H}_1^{(inter)}$ chooses all interactions with indexes $(m = \pm 1,m^\prime = \pm 1)$ and $(m = 0 ,m^\prime = 0)$. 
\begin{figure}[h]
	\includegraphics[width= 0.50\textwidth]{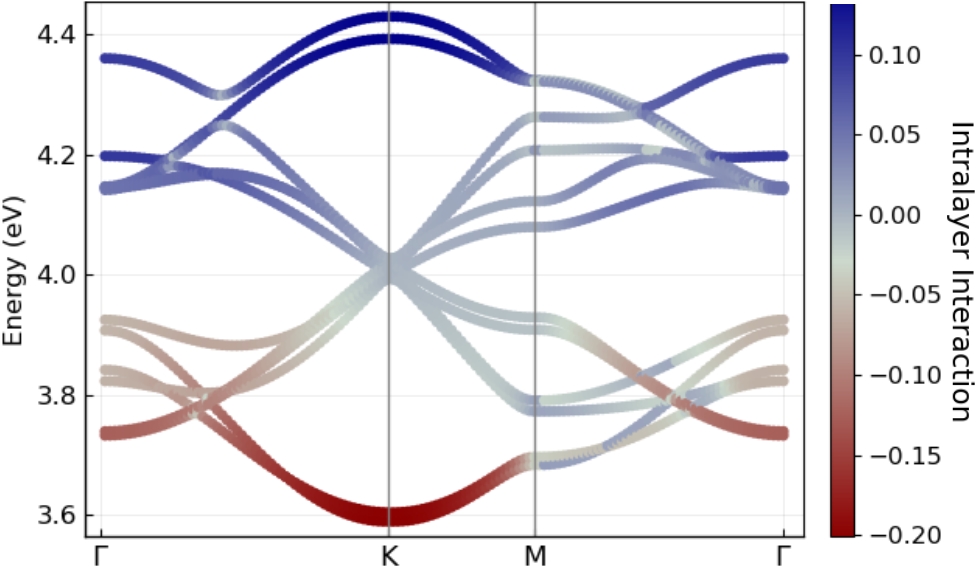}
	\caption{  Percentage of $n_s$ contributed from intralyer interactions along high-symmetry points,$\left< \mathbf{H}^{(intra)}\right>_s/n_s$ , complimentary to Figs.~\ref{fig:5}.   }
	\label{fig:app3}
\end{figure}
On the other hand $\mathbf{H}_2^{(inter)}$ includes all interactions of the form $ (m = 0, m^\prime \pm1)$. A schematic of these types of interactions is shown in Fig.~\ref{fig:app2}.
Finally, as a compliment to section the decomposition of the interactions in the vertically stacked layers, we present the intralayer interaction along the high-symmetry points.

\bibliography{BilayersSR}
\end{document}